\documentclass[%
 reprint,
 amsmath,amssymb,
 aps,
 prc,
floatfix,
]{revtex4-2}
\usepackage{longtable}

\usepackage{changes}
\usepackage[para,online,flushleft]{threeparttable}
\usepackage{graphicx}  % needed for figures
\usepackage{dcolumn}   % needed for some tables
\usepackage{bm}        % for math
\usepackage{amsmath,amsfonts} % for math (amssymb comes from the class options)
\usepackage{braket}    % for QM notation
\usepackage{booktabs}
\usepackage{multirow}
\usepackage{adjustbox}
\usepackage[T1]{fontenc}
\usepackage[utf8]{inputenc}
\DeclareUnicodeCharacter{039B}{\ensuremath{\Lambda}}
\usepackage[normalem]{ulem}
\usepackage{color}
\usepackage{dsfont}
\usepackage{tikz}
\usepackage{pgfplots}
\usepgfplotslibrary{fillbetween}
\usetikzlibrary {arrows.meta}
\usepackage[section]{placeins}
\usepackage{enumitem}
\usepackage{capt-of}
\usepackage[colorlinks=true, linkcolor=blue, citecolor=blue, urlcolor=blue]{hyperref}
\usepgfplotslibrary{groupplots}
\pgfplotsset{compat=newest}
\usepgfplotslibrary{units}
\usetikzlibrary{patterns}

\definecolor{darkgreen}{rgb}{0.0,0.5,0.0}

\definecolor{revisiongreen}{rgb}{0.0,0.5,0.0}
\definechangesauthor[name={Revision}, color=revisiongreen]{rev}

\begin{document}

\title{Constraining hyperonic relativistic mean-field models with \\ rapidly rotating neutron stars}
\author{Gihwan Nam}
\email{namgh@yonsei.ac.kr}
\affiliation{Department of Physics, Yonsei University, Seoul 03722, South Korea}
\author{Yeunhwan Lim}
\email{ylim@yonsei.ac.kr}
\affiliation{Department of Physics, Yonsei University, Seoul 03722, South Korea}
\author{Prashant Thakur}
\email{prashant@yonsei.ac.kr}
\affiliation{Department of Physics, Yonsei University, Seoul 03722, South Korea}
\author{Jeremy~W. Holt}
 \email{holt@physics.tamu.edu}
\affiliation{Cyclotron Institute, Texas A\&M University, College Station, TX 77843, USA}
\affiliation{Department of Physics and Astronomy, Texas A\&M University, College Station, TX 77843, USA }
%%\vskip 0.25cm
\date{\today}

\begin{abstract}
Motivated by the recent mass measurement of the black-widow pulsar
PSR~J0952$-$0607 with $M=2.35\pm0.11\,M_\odot$,
we investigate how the masses of heavy, rapidly rotating
millisecond pulsars can be used to constrain
relativistic mean-field (RMF) models containing hyperonic
degrees of freedom. %and rapid rotation treated consistently within one framework.
%Starting from a previously derived empirical relation for the maximum mass of nonrotating nucleonic neutron stars within the RMF framework, we construct new equations of state (EOSs) and analyze the resulting stellar properties.
In our approach, hyperons are incorporated following the spin-flavor SU(6) symmetry scheme for the vector-meson couplings. We find that increasing the nonlinear $\omega$-meson vector self-coupling parameter $\zeta$ suppresses the hyperon fraction and can alter the onset ordering of the $\Sigma^-$ and $\Xi^-$ hyperons.
By computing rotating neutron-star configurations at the observed spin
frequency $707\,\mathrm{Hz}$ of PSR~J0952$-$0607, we identify RMF models compatible with this pulsar's observed lower-mass bound. Using an empirical relation for the maximum neutron star mass, the PSR~J0952$-$0607 observational contraint is mapped onto the allowed RMF parameter space in $n_0$, $m^\ast$, and $\zeta$.
\end{abstract}
\maketitle

\section{Introduction}
Massive stars undergo supernova explosions after exhausting their
nuclear fuel, leaving behind compact remnants known as neutron
stars~\cite{Prialnik2010}.
%Because stellar nucleosynthesis terminates at iron, neutron stars are commonly described as cold-catalyzed matter in their ground state~\cite{Haensel2006}.
Their birth masses and radii depend on the
progenitor properties and on the dynamics of the core-collapse
supernova~\cite{Haensel2006}, while the stars themselves are expected
to lie on a single mass--radius curve determined by the equation of
state~\cite{lattimer2012}.
As the neutron-star mass increases, matter in the
core is compressed to several times nuclear saturation density and
becomes highly neutron rich~\cite{lim19e,lim2024}.
Neutron stars therefore provide a unique probe of nuclear interactions in dense,
highly isospin-asymmetric matter~\cite{Glendenning2000}.
Accurate neutron-star mass and radius measurements
remain challenging, particularly for radii. Nevertheless,
accurate radius measurements of canonical neutron stars with masses
around $1.4\,M_\odot$ can provide direct constraints on the equation
of state~\cite{Steiner05}.
This is closely connected to the density dependence of the nuclear
symmetry energy, especially its slope parameter $L$, which has strong
implications for both neutron-star radii and neutron-skin measurements
such as the PREX experiment for
$^{208}\mathrm{Pb}$~\cite{lattimer13,Horowitz2001,Fattoyev2018,Adhikari2021,Reed2021}.

In recent years, several massive neutron stars have been
observed~\cite{Demorest2010,Cromartie2020,Fonseca2021}.
However, although various equations of state have been proposed to
satisfy the observational constraints, the implications of these
observations for the nuclear equation of state have not yet been fully
understood in a systematic
way~\cite{Virender2022,Fan2022,Miyatsu2023}.
Motivated by this, empirical relations \cite{Nam2026} within the framework of
relativistic mean-field (RMF) theory have been investigated, showing that
the maximum mass of neutron stars can be described in a simple form in
terms of the saturation density $(n_0)$, the dimensionless nucleon
effective mass at saturation
$m^\ast\equiv m_N^\ast(n_0)/m$, and the strength of the nonlinear
self-coupling $(\zeta)$ of the omega field. Throughout
this work, $m^\ast$ denotes this dimensionless saturation value, while
$m_N^\ast(n)$ denotes the dimensionful Dirac effective mass at density
$n$.
This suggests that observations of massive neutron stars can provide
constraints on these saturation properties and on $\zeta$.

The core contains most of the neutron star's mass and is expected to exhibit a rich variety of dense-matter phenomena.
The outer core is generally believed to consist mainly of nucleons,
electrons, and muons, where neutrons and protons may form superfluid
and superconducting phases,
respectively~\cite{Lattimer2004,Sheternin2011,Lim2021p,ma26}.
At higher densities in the inner core, additional exotic degrees of
freedom such as hyperons, meson condensates, or deconfined quark
matter may appear~\cite{thorsson94,glen98,alford05,weber05}.
Among these possibilities, the appearance of hyperons is considered
particularly plausible~\cite{Glendenning1982,Glendenning1991}.
Owing to the relatively small mass difference between hyperons and
nucleons, the appearance of hyperons becomes energetically favorable
as the neutron chemical potential increases with density.
Consequently, hyperons may emerge in chemical equilibrium at densities
of a few times the nuclear saturation density
~\cite{Schaffner2008,Vidana2016,lim18h}.
Although hyperons are plausible constituents of
dense neutron-star matter, their appearance generally softens the EOS
and lowers the maximum mass relative to that of the corresponding
nucleonic EOS.
This is the well-known hyperon puzzle posed by observations of massive
neutron stars~\cite{weissenborn12b,Lonardoni:2013rm}. In particular, the millisecond
pulsar PSR~J0952$-$0607 provides one of the most stringent constraints
among the currently known massive neutron stars~\cite{Bassa2017,Romani2022}, and
its mass has recently been tightened to
$2.35\pm0.11\,M_\odot$~\cite{Romani2026}, which is the immediate
motivation for the present analysis.
In addition, PSR~J0952$-$0607 has the rotational frequency $707\,\mathrm{Hz}$, fast enough that its
maximum mass has to be evaluated on the rotating
sequence~\cite{Komatsu1989,Stergioulas1995,Rezzolla:2018jee}, rather than in the nonrotating limit.

In the present work, we investigate how RMF models are constrained when hyperonic degrees of freedom and rotational effects associated
with millisecond pulsars are taken into account, based on the
previously derived empirical relation~\cite{Nam2026} for the maximum mass of spherical nucleon-only neutron stars within that framework.
Our emphasis is on the construction itself rather than on advocating
one particular numerical bound: given a threshold mass and a spin
frequency, the empirical relation converts them into an explicit
allowed region in $(n_0,m^\ast,\zeta)$ parameter space.
Because pulsar mass measurements continue to be revised, we treat the
threshold as an adjustable input and explicitly examine how the allowed parameter space depends on its value.
The paper is organized as follows.
In section \ref{sec:formalism}, we describe the RMF formalism employed
in this work.
In section \ref{sec:results}, we present and discuss the results. We end with a summary and outlook.

\section{Formalism}\label{sec:formalism}

\subsection{RMF Lagrangian and coupling scheme}

Relativistic mean-field theory is widely used in neutron star modeling because it can consistently describe both finite nuclei (in neutron star crusts) and bulk nuclear matter (in neutron star cores). In addition, the formalism is relatively straightforward to
implement\,\cite{walecka1974,Reinhard1989,Dutra2014,Serot1992,Serot1997}.
The nucleonic Lagrangian density employed in this work is given by
\begin{align} \label{eq:L_N}
\mathcal{L}_{N,\sigma \omega \rho}
&= \bar\psi\left[\gamma^\mu\bigl(i\partial_\mu - g_\omega \omega_\mu
-\frac{g_\rho}{2}\bm{\tau}\cdot \bm{\rho}_\mu\bigr)
-(m-g_\sigma \sigma)\right]\psi \nonumber \\
&\quad
+ \tfrac12\,\partial_\mu\sigma\,\partial^\mu\sigma
- \tfrac12\,m_\sigma^2\,\sigma^2
- \tfrac{\kappa}{3}(g_\sigma \sigma)^3
- \tfrac{\lambda}{4}(g_\sigma \sigma)^4 \nonumber \\
&\quad
- \tfrac14\,W_{\mu\nu}\,W^{\mu\nu}
+ \tfrac12\,m_\omega^2\,\omega_\mu\,\omega^\mu
+ \tfrac{\zeta}{4}
  \bigl(g_\omega^2\,\omega_\mu\omega^\mu\bigr)^2 \nonumber\\
&\quad
- \tfrac14\,\bm{R}_{\mu\nu}\cdot\bm{R}^{\mu\nu}
+ \tfrac12\,m_\rho^2\,\bm{\rho}_\mu\cdot\bm{\rho}^{\mu}
\nonumber\\
&\quad
+ \Lambda_{s1}(g_\sigma\sigma)
  \bigl(g_\rho^2\,\bm{\rho}_\mu\cdot\bm{\rho}^{\mu}\bigr)
\nonumber\\
&\quad
+ \Lambda_{s2}(g_\sigma\sigma)^2
  \bigl(g_\rho^2\,\bm{\rho}_\mu\cdot\bm{\rho}^{\mu}\bigr).
\end{align}
Detailed descriptions of the RMF framework and its formalism can be
found in
Refs.~\cite{Fattoyev2010,Dutra2014,Malik:2023mnx,Tolos:2017lgv}.
Here, $\psi$ denotes the nucleon field with bare mass $m$, interacting
through the meson fields $\sigma$, $\omega_\mu$, and $\bm{\rho}_\mu$,
which represent the scalar, isoscalar-vector, and isovector-vector
channels, respectively.
The corresponding coupling constants are $g_\sigma$, $g_\omega$, and
$g_\rho$, and $\bm{\tau}$ is the nucleon isospin operator.
The meson masses are denoted by $m_\sigma$, $m_\omega$, and $m_\rho$.
The field-strength tensors of the $\omega$ and $\rho$ mesons are
\begin{align*}
W_{\mu\nu}
&=\partial_\mu\omega_\nu-\partial_\nu\omega_\mu,\\
\bm{R}_{\mu\nu}
&=\partial_\mu\bm{\rho}_\nu-\partial_\nu\bm{\rho}_\mu.
\end{align*}
The parameters $\kappa$, $\lambda$, and $\zeta$ govern the nonlinear
meson self-interactions~\cite{boguta1977,Muller1996}.
The additional couplings $\Lambda_{s1}$ and $\Lambda_{s2}$ describe
$\sigma$--$\rho$ mixed interactions and provide additional flexibility
in controlling the density dependence of the symmetry energy.
Compared with the commonly used $\omega$--$\rho$ mixing scheme, the
$\sigma$--$\rho$ interactions allow a convenient adjustment to the
pure-neutron-matter
constraints~\cite{Horowitz2001,Todd-Rutel2005,Fattoyev2010}.

The lepton fields $\psi_l$ ($l=e,\mu$), with masses $m_l$, are treated
as noninteracting free Fermi gases,
\begin{equation}\label{eq:L_lep}
\mathcal{L}_{l}
=
\sum_{l=e,\mu}
\bar\psi_l
\left(i\gamma^\mu\partial_\mu-m_l\right)
\psi_l .
\end{equation}
The baryonic sector is extended to include the hyperon octet through
\begin{align}\label{eq:L_Y}
\mathcal{L}_{Y,\sigma\omega\rho\phi}
&= \sum_Y \bar\psi_Y
\Bigl[
\gamma^\mu\bigl(
i\partial_\mu
-g_{\omega Y}\omega_\mu
-\tfrac{g_\rho}{2}\bm{\tau}_Y\cdot\bm{\rho}_\mu
\nonumber\\
&\hspace{2.3cm}
-g_{\phi Y}\phi_\mu
\bigr)
-(m_Y-g_{\sigma Y}\sigma)
\Bigr]\psi_Y
\nonumber\\
&\quad
-\frac14 F_{\mu\nu}^{(\phi)}F^{(\phi)\mu\nu}
+\frac12 m_\phi^2\phi_\mu\phi^\mu
.
\end{align}
Here, $\psi_Y$ denotes a hyperon field with mass $m_Y$, and
$Y=\{\Lambda,\Sigma^-,\Sigma^0,\Sigma^+,\Xi^-,\Xi^0\}$.
As in the nucleon sector, we assume isospin symmetry and use a common
mass within each hyperon multiplet: $m_\Lambda=1115\,\mathrm{MeV}$,
$m_\Sigma=1193\,\mathrm{MeV}$, and $m_\Xi=1318\,\mathrm{MeV}$.
The field-strength tensor of the hidden-strangeness vector meson is
$F_{\mu\nu}^{(\phi)}=\partial_\mu\phi_\nu-\partial_\nu\phi_\mu$.
The hyperons interact with the scalar meson $\sigma$, the vector
mesons $\omega_\mu$ and $\phi_\mu$, and the isovector-vector meson
$\bm{\rho}_\mu$.
The $\phi$ meson is composed of strange quarks and does not couple to
nucleons.
It provides additional repulsion among hyperons and thereby alleviates
the softening of the EOS induced by the appearance of hyperons
\cite{Weissenborn2012}.
The corresponding hyperon--meson coupling constants are $g_{\sigma
Y}$, $g_{\omega Y}$, and $g_{\phi Y}$; the isovector coupling $g_\rho$ is
common to all baryons, the species dependence being carried by the isospin
projection $I_{3B}$.

The vector-meson couplings are determined according to the SU(6)
spin--flavor symmetry of the quark model
\cite{DOVER1984,SCHAFFNER1994}.
The resulting coupling relations are
\begin{equation}
g_{\omega\Lambda}=g_{\omega\Sigma}=\frac{2}{3}g_\omega,
\qquad
g_{\omega\Xi}=\frac{1}{3}g_\omega,
\label{eq:omega_Y_couplings}
\end{equation}
\begin{equation}
g_{\phi\Lambda}=g_{\phi\Sigma}
=-\frac{\sqrt{2}}{3}g_\omega,
\qquad
g_{\phi\Xi}=-\frac{2\sqrt{2}}{3}g_\omega,
\label{eq:phi_Y_couplings}
\end{equation}
The scalar couplings $g_{\sigma Y}$ are determined from the hyperon
potential depths in symmetric nuclear matter at saturation density,
\begin{equation}
U_Y^{(N)}(n_0)
=g_{\omega Y}\omega_0-g_{\sigma Y}\sigma_0 .
\label{eq:hyperon_potential}
\end{equation}
We adopt $U_\Lambda^{(N)}=-30\,\mathrm{MeV}$,
$U_\Sigma^{(N)}=+30\,\mathrm{MeV}$, and
$U_\Xi^{(N)}=-18\,\mathrm{MeV}$ \cite{SchaffnerBielich2000}.
The $\Lambda$ potential is relatively well constrained by hypernuclear
experiments, whereas the $\Sigma$ and $\Xi$ potentials remain less
certain \cite{Hashimoto2006,Haidenbauer2020,Haidenbauer2023}.
Current studies generally favor a repulsive $\Sigma$ potential and a
weakly attractive $\Xi$ potential \cite{GalAvraham2016,Vidana2018}.
The effects of varying these potential depths on neutron-star
sequences have been investigated in
Refs.~\cite{Bhowmick2014,Weissenborn2012}.

\subsection{Uniform matter in the mean-field approximation}

To construct neutron stars, we first consider cold, catalyzed, and spatially uniform matter.
In the mean-field approximation, the meson operators are replaced by
their uniform expectation values,
\begin{align}
\langle\sigma\rangle&=\sigma,
&
\langle\omega^\mu\rangle&=\omega_0,
\nonumber\\
\langle\bm{\rho}^\mu\rangle&=\rho_{03},
&
\langle\phi^\mu\rangle&=\phi_0 .
\label{eq:uniform_mean_fields}
\end{align}
Only the following components are nonzero: $\omega_0$ and $\phi_0$ are the time
components, while $\rho_{03}$ is the time component along the third
direction in isospin space.
Hereafter, $N\in\{n,p\}$ denotes a nucleon, $Y$ denotes a hyperon, and
$B\in\{N,Y\}$ denotes a generic baryon.
Their Dirac effective masses are
\begin{equation}
m_N^\ast=m-g_\sigma\sigma,
\qquad
m_Y^\ast=m_Y-g_{\sigma Y}\sigma .
\label{eq:effective_masses}
\end{equation}
For a spin-$1/2$ baryon with Fermi momentum $k_B$, the vector and
scalar densities are
\begin{align}
n_B&=\frac{k_B^3}{3\pi^2},
\label{eq:vector_density}\\
n_s^B
&=\frac{1}{\pi^2}
\int_0^{k_B}
\frac{m_B^*k^2\,dk}{\sqrt{k^2+m_B^{*2}}}
,
\label{eq:scalar_density}
\end{align}
Variation of the mean-field Lagrangian with respect to the meson
fields yields the following coupled field equations:
\begin{align}
&m_\sigma^2\sigma
+\kappa g_\sigma(g_\sigma\sigma)^2
+\lambda g_\sigma(g_\sigma\sigma)^3
\nonumber\\
&\qquad
-g_\sigma(g_\rho\rho_{03})^2
\left[
\Lambda_{s1}
+2\Lambda_{s2}(g_\sigma\sigma)
\right]
=
g_\sigma(n_s^n+n_s^p)
\nonumber\\
&\qquad
+\sum_Y g_{\sigma Y}n_s^Y ,
\label{eq:sigma_field}
\\
&m_\omega^2\omega_0
+\zeta g_\omega^4\omega_0^3
=
g_\omega(n_n+n_p)
+\sum_Y g_{\omega Y}n_Y ,
\label{eq:omega_field}
\\
&\left\{
m_\rho^2
+2g_\rho^2
\left[
\Lambda_{s1}(g_\sigma\sigma)
+\Lambda_{s2}(g_\sigma\sigma)^2
\right]
\right\}\rho_{03}
\nonumber\\
&\qquad
=
\frac{g_\rho}{2}(n_p-n_n)
+g_\rho\sum_Y I_{3Y}n_Y ,
\label{eq:rho_field}
\\
&m_\phi^2\phi_0
=
\sum_Yg_{\phi Y}n_Y .
\label{eq:phi_field}
\end{align}
For the self-consistent meson fields, the nucleon, hyperon, and lepton
chemical potentials are
\begin{align}
\mu_N
&=
\sqrt{k_N^2+m_N^{*2}}
+g_\omega\omega_0
+g_\rho I_{3N}\rho_{03},
\label{eq:mu_N}\\
\mu_Y
&=
\sqrt{k_Y^2+m_Y^{*2}}
+g_{\omega Y}\omega_0
+g_\rho I_{3Y}\rho_{03}
+g_{\phi Y}\phi_0,
\label{eq:mu_Y}\\
\mu_l
&=
\sqrt{k_l^2+m_l^2}.
\label{eq:mu_lep}
\end{align}
The third components of nucleon isospin are
\begin{equation}
I_{3n}=-\frac{1}{2},\qquad I_{3p}=+\frac{1}{2},
\end{equation}
and those of the hyperons are
\begin{align}
I_{3\Lambda}=0,\quad
I_{3\Sigma^-}=-1,\quad
I_{3\Sigma^0}=0,
\nonumber\\
I_{3\Sigma^+}=+1,\quad
I_{3\Xi^-}=-\frac{1}{2},\quad
I_{3\Xi^0}=+\frac{1}{2} .
\label{eq:hyperon_isospin}
\end{align}

Neutrino-transparent matter in $\beta$ equilibrium satisfies
\begin{align}
\mu_\Lambda&=\mu_n,
&
\mu_{\Sigma^-}&=\mu_n+\mu_e,
&
\mu_{\Sigma^0}&=\mu_n,
\nonumber\\
\mu_{\Sigma^+}&=\mu_n-\mu_e,
&
\mu_{\Xi^-}&=\mu_n+\mu_e,
&
\mu_{\Xi^0}&=\mu_n,
\label{eq:beta_eq}
\end{align}
together with $\mu_\mu=\mu_e$ when muons are present.
These relations can be written compactly as $\mu_B=\mu_n-q_B\mu_e$,
where $q_B$ is the electric charge of baryon $B$ in units of the
proton charge.
Charge neutrality requires
\begin{equation}
n_p+n_{\Sigma^+}
-n_{\Sigma^-}-n_{\Xi^-}
-n_e-n_\mu=0 ,
\label{eq:charge_neutrality}
\end{equation}
and the total baryon density is
\begin{equation}
n=n_n+n_p+\sum_Yn_Y .
\label{eq:total_baryon_density}
\end{equation}
The meson-field equations, $\beta$-equilibrium conditions, charge
neutrality, and the fixed total density are solved simultaneously.

In $\beta$-equilibrated matter, hyperon thresholds are
governed first by the mass hierarchy among the strange members of the
baryon octet.
The $\Lambda$, an isospin singlet and the lightest hyperon, therefore
generally appears first.
Within the $\Sigma$ triplet and $\Xi$ doublet, however, approximate
$SU(2)$ isospin symmetry keeps the members of each multiplet nearly
degenerate in mass, so their relative onset ordering is determined
mainly by the competition between charge neutrality and the isospin
interaction.
Charge neutrality favors negatively charged members, whereas the
isospin interaction may favor members with larger electric charge
\cite{Glendenning2000,Haensel2006}.
In most cases, the charge-neutrality effect dominates.
As negatively charged hyperons replace leptons, however, the resulting
decrease in the lepton chemical potential weakens this preference,
bringing the onset densities of members within the same multiplet
closer together.

\subsection{Equation of state}

For later use, the kinetic contributions of baryon $B$ and lepton $l$
to the energy density and pressure are defined as
\begin{align}
\varepsilon_B^{\mathrm{kin}}
&=\frac{1}{\pi^2}\int_0^{k_B}
k^2\sqrt{k^2+m_B^{*2}}\,dk,
\label{eq:baryon_kinetic_energy}\\
P_B^{\mathrm{kin}}
&=\frac{1}{3\pi^2}\int_0^{k_B}
\frac{k^4\,dk}{\sqrt{k^2+m_B^{*2}}},
\label{eq:baryon_kinetic_pressure}\\
\varepsilon_l^{\mathrm{kin}}
&=\frac{1}{\pi^2}\int_0^{k_l}
k^2\sqrt{k^2+m_l^2}\,dk,
\label{eq:lepton_kinetic_energy}\\
P_l^{\mathrm{kin}}
&=\frac{1}{3\pi^2}\int_0^{k_l}
\frac{k^4\,dk}{\sqrt{k^2+m_l^2}}.
\label{eq:lepton_kinetic_pressure}
\end{align}
The total energy density obtained from the energy--momentum tensor is given by
\begin{align}
\varepsilon
&=
\sum_B\varepsilon_B^{\mathrm{kin}}
+\sum_{l=e,\mu}\varepsilon_l^{\mathrm{kin}}
\nonumber\\
&\quad
+\sum_B
\left(
g_{\omega B}\omega_0
+g_{\phi B}\phi_0
+g_\rho I_{3B}\rho_{03}
\right)n_B
\nonumber\\
&\quad
+\frac12m_\sigma^2\sigma^2
+\frac{\kappa}{3}(g_\sigma\sigma)^3
+\frac{\lambda}{4}(g_\sigma\sigma)^4
\nonumber\\
&\quad
-\frac12m_\omega^2\omega_0^2
-\frac{\zeta}{4}(g_\omega\omega_0)^4
-\frac12m_\rho^2\rho_{03}^2
-\frac12m_\phi^2\phi_0^2
\nonumber\\
&\quad
-\Lambda_{s1}(g_\sigma\sigma)(g_\rho\rho_{03})^2
-\Lambda_{s2}(g_\sigma\sigma)^2(g_\rho\rho_{03})^2 ,
\label{eq:energy_density}
\end{align}
where $g_{\omega N}=g_\omega$ and $g_{\phi N}=0$,
so that the $\phi$ contribution is carried by the hyperons alone.
The corresponding pressure is
\begin{align}
P
&=
\sum_BP_B^{\mathrm{kin}}
+\sum_{l=e,\mu}P_l^{\mathrm{kin}}
-\frac12m_\sigma^2\sigma^2
-\frac{\kappa}{3}(g_\sigma\sigma)^3
\nonumber\\
&\quad
-\frac{\lambda}{4}(g_\sigma\sigma)^4
+\frac12m_\omega^2\omega_0^2
+\frac{\zeta}{4}(g_\omega\omega_0)^4
\nonumber\\
&\quad
+\frac12m_\rho^2\rho_{03}^2
+\frac12m_\phi^2\phi_0^2
\nonumber\\
&\quad
+\Lambda_{s1}(g_\sigma\sigma)(g_\rho\rho_{03})^2
+\Lambda_{s2}(g_\sigma\sigma)^2(g_\rho\rho_{03})^2 .
\label{eq:pressure}
\end{align}
These expressions satisfy the zero-temperature thermodynamic identity,
\begin{equation}
P=\sum_B\mu_Bn_B+\sum_{l=e,\mu}\mu_ln_l-\varepsilon .
\label{eq:thermodynamic_identity}
\end{equation}
Solving the coupled equations for a given $n$
yields the EOS
$P(\varepsilon)$ used in the stellar-structure calculations.

\subsection{Uniformly rotating configurations}

To compute equilibrium configurations of uniformly rotating neutron
stars, we use the publicly available \texttt{RNS} code
\cite{Stergioulas1995}.
The stellar matter is treated as a barotropic perfect fluid,
\begin{equation*}
T^{\mu\nu}=(\varepsilon+P)u^\mu u^\nu+Pg^{\mu\nu},
\qquad
u^\mu=u^t(1,0,0,\Omega),
\end{equation*}
where the angular velocity $\Omega$ is uniform throughout the star.
The code constructs stationary, axisymmetric models by solving the
Einstein equations together with hydrostationary equilibrium for the
line element \cite{Komatsu1989}
\begin{equation}
\begin{aligned}
ds^2={}&-e^{\gamma+\rho}dt^2
+e^{2\alpha}(dr^2+r^2d\theta^2)
\\
&+e^{\gamma-\rho}r^2\sin^2\theta
(d\phi-\omega dt)^2 .
\end{aligned}
\label{eq:rns_metric}
\end{equation}
Stationarity and axisymmetry imply that the four metric potentials
$\gamma$, $\rho$, $\alpha$, and $\omega$ depend only on $r$ and
$\theta$.
For a specified EOS, central energy density, and rotation rate, these
potentials and the fluid distribution are determined self-consistently.
The lapse is $e^{(\gamma+\rho)/2}$, whereas the proper circumferential
radius at $(r,\theta)$ is $e^{(\gamma-\rho)/2}r\sin\theta$.
The potential $\alpha$ determines proper distances in the meridional
$(r,\theta)$ plane.
Finally, $\omega$ is the angular velocity of local inertial-frame
dragging: a zero-angular-momentum observer satisfies
$d\phi/dt=\omega$. 
%\tcm{YH:
%Is $\Omega$ is the stellar matter, and $\omega$ for fluid inside the stellar matter?? What is the exact difference between $\Omega$ and $\omega$?}
%\tcb{GH: $\Omega$ belongs to the matter. It is the rate $d\phi/dt$ at which a fluid element circulates, in the coordinate time $t$ of a distant observer. Under rigid rotation it is a single constant and it is an input we impose.
%$\omega$ belongs to the spacetime, not to any fluid. It is the rate at which a local inertial frame is itself dargged around. It is a metric function; it varies with position, it is solved for, and it stays nonzero in the vacuum outside the star, where there is no fluid at all. It dies as $\sim 1/r^3$. I added new sentences.}
It must be distinguished from the uniform angular velocity $\Omega$ of
the stellar fluid.
Whereas $\Omega$ is a single constant characterizing the matter, $\omega=\omega(r,\theta)$ is a
metric field that remains nonzero in the vacuum exterior, falling off as $2J/r^3$. The fluid
rotates relative to the local inertial frame at the rate $\Omega-\omega$, and it is this
combination that enters the hydrostationary equilibrium through the proper velocity$v=(\Omega-\omega)\,r\sin\theta\,e^{-\rho}$ measured by a zero-angular-momentum observer.
In the nonrotating limit, $\omega=0$ and the remaining potentials
depend only on $r$, so
Eq.~\eqref{eq:rns_metric} reduces to the static, spherically symmetric
metric in isotropic coordinates.
Note that the symbol $\rho$ in Eq.~\eqref{eq:rns_metric} denotes a metric
function and is unrelated to the $\rho$ meson.
For each EOS we construct a sequence of rigidly rotating models at
the fixed spin frequency $f=\Omega/2\pi=707\,\mathrm{Hz}$ of
PSR~J0952$-$0607 by varying the central energy density, and identify
the maximum gravitational mass $M_{Y,707}$ along the sequence.

The size of the rotational correction is set by the ratio of the
spin to the mass-shedding frequency $f_{\mathrm{K}}$, which for massive
neutron stars is of order $1.2$--$1.5\,\mathrm{kHz}$.
In the slow-rotation regime the increase of the maximum mass grows
approximately as $(f/f_{\mathrm{K}})^{2}$~\cite{Hartle1967} and
steepens as the mass-shedding limit is approached, where it reaches
about $20\%$ for tabulated hadronic equations of
state~\cite{Cipolletta2015,Breu2016}. A broader range of maximum-mass enhancements is found for a more generic set of equations of state \cite{Musolino2024} and models with an abrupt
first-order softening~\cite{Bozzola2019}.
Spins of a few hundred hertz, such as the $346\,\mathrm{Hz}$ of
PSR~J0740+6620~\cite{Fonseca2021}, give a correction some four times
smaller, negligible against the quoted mass uncertainties.
The maximum mass used with the fastest millisecond pulsars is therefore
taken from the rotating sequence, as here, or from universal relations
between rotating and nonrotating masses~\cite{Konstantinou2022}.

Throughout, we use different notations for the maximum mass depending
on the composition and rotational state of the neutron star.
The maximum masses of the nucleon-only non-rotating, hyperonic
non-rotating, hyperonic $707\,\mathrm{Hz}$ rotating, and hyperonic
Keplerian configurations are denoted by $M_{N,\mathrm{TOV}}$,
$M_{Y,\mathrm{TOV}}$, $M_{Y,707}$, and $M_{Y,\mathrm{Kep}}$,
respectively.

\section{Results}\label{sec:results}
% tall 3-panel float: "!" is needed to override \topfraction
\begin{figure}[!t]
    \centering
    \includegraphics[width=1.0\linewidth]{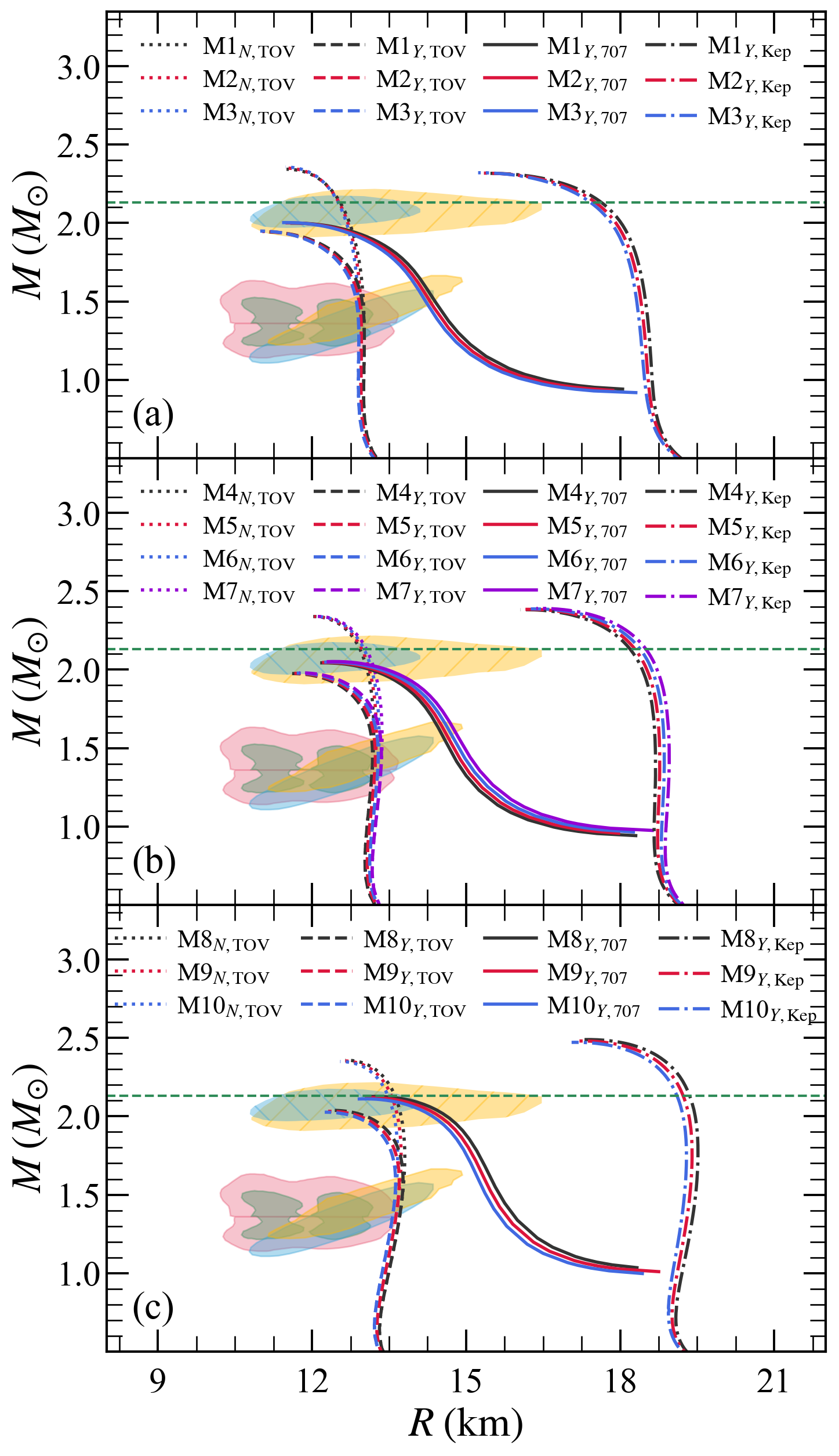}
\caption{Mass--radius relations for the RMF model sets
M1--M10\,\cite{Nam2026}, grouped by the vector self-coupling parameter
$\zeta$: from top to bottom, (a) M1--M3 with $\zeta=0$, (b) M4--M7
with $\zeta=0.001$, and (c) M8--M10 with $\zeta=0.002$.
Dotted curves denote nucleon-only TOV solutions, dashed curves
represent hyperonic TOV configurations, solid curves correspond to
rotating stars at a fixed spin frequency of $707\,\mathrm{Hz}$, and
dash-dotted curves indicate the Kepler (mass-shedding) limit.
The horizontal green dashed line shows the $2\sigma$ lower mass bound
inferred from PSR~J0952$-$0607, $2.13\,M_\odot$ \cite{Romani2026}.
The shaded regions show observational constraints: the green and red
regions represent the 50\% and 90\% credible regions from GW170817
\cite{abbott2018b}, the yellow and blue hatched regions indicate the
68\% credible regions from NICER analyses of PSR~J0740+6620
by Miller et al.\ and Riley et al.\ \cite{Miller2021,Riley2021}, and
the yellow and blue shaded regions correspond to the 68\% credible regions 
from PSR~J0030+0451 by Miller et al.\ and Riley et al.,
respectively \cite{Miller2019,Riley2019}.} 
\label{fig:M1-10}
\end{figure}

\begin{figure}[t]
    \centering
    \includegraphics[width=1.0\linewidth]{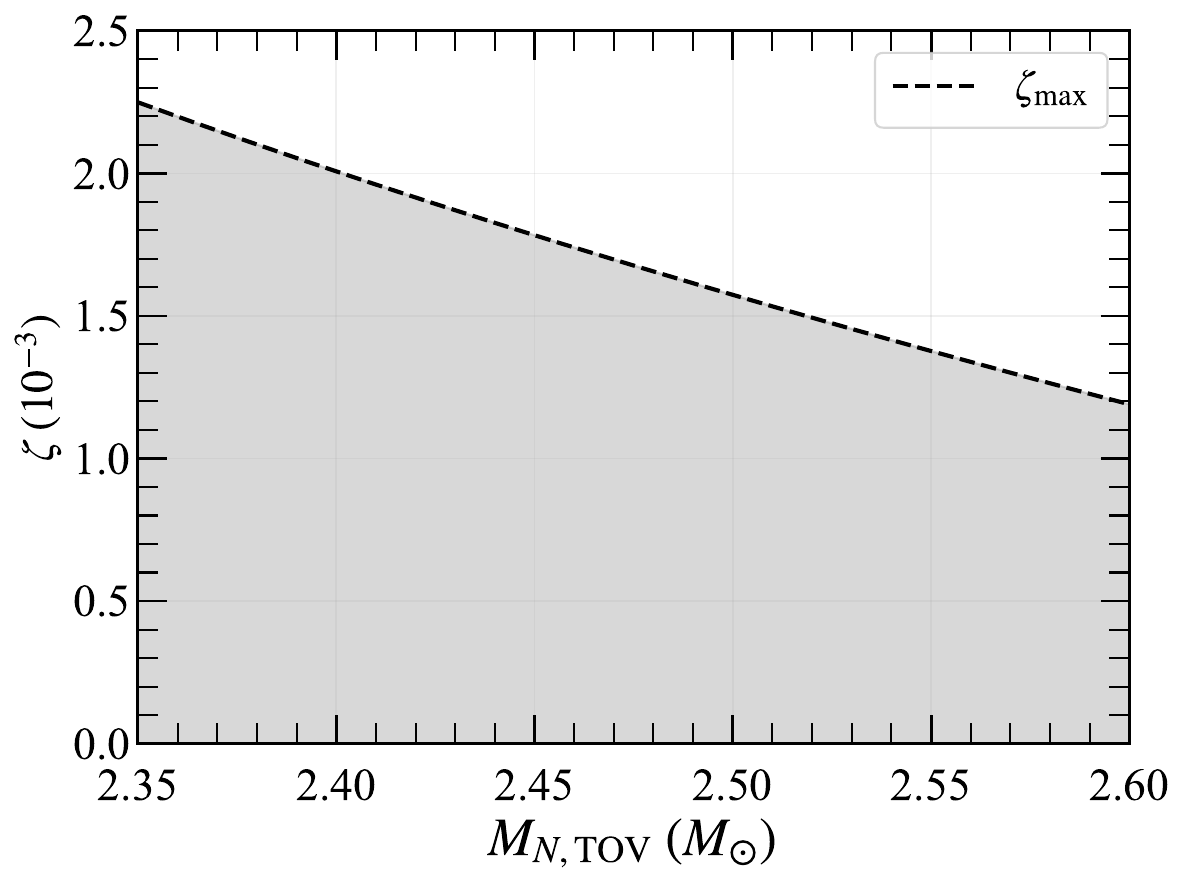}
    \caption{Upper limit $\zeta_{\max}$ as a function of the nucleonic
    TOV maximum mass $M_{N,\mathrm{TOV}}$, evaluated at
    $n_0=0.145\,\mathrm{fm^{-3}}$ and $m^\ast=0.54$.
    The shaded region denotes $0\leq\zeta\leq\zeta_{\max}$.}
    \label{fig:ZetaRange}
\end{figure}

The recently tightened mass measurement of PSR~J0952$-$0607,
$2.35\pm0.11\,M_\odot$~\cite{Romani2026}, provides a stringent
observational lower bound on the maximum mass that a viable equation
of state must support. We adopt the $2\sigma$ lower edge of this
measurement, $2.13\,M_\odot$, rather than the corresponding $1\sigma$
value of $2.24\,M_\odot$. This conservative choice reflects the fact
that the mass estimate relies on optical light-curve and radial-velocity
modeling of an irradiated companion, which is subject to modeling
systematics. A cautionary example is PSR~J0348+0432, whose mass was
originally inferred to be $2.01\pm0.04\,M_\odot$ from optical
observations~\cite{Antoniadis2013}, but was subsequently revised
downward to $1.806\pm0.037\,M_\odot$ after substantially improved
pulsar-timing constraints became available~\cite{Saffer2025}. We
therefore use $2.13\,M_\odot$ as the fiducial observational lower bound
rather than treating the $1\sigma$ edge as a hard constraint on the EOS.

Figure~\ref{fig:M1-10} shows the mass--radius relations obtained from
the previously constructed RMF model sets M1--M10.
These models were obtained such that the nucleon-only EOS reproduces a
maximum mass of $M_{\max}=2.35\,M_\odot$ in the nonrotating (TOV)
limit, chosen in light of the mass measurement of PSR~J0952$-$0607.
The models are grouped according to the vector self-coupling parameter
$\zeta$, and each group is shown in a separate panel: (a) M1--M3 with
$\zeta=0$, (b) M4--M7 with $\zeta=0.001$, and (c) M8--M10 with
$\zeta=0.002$.
Within each group, the saturation properties $n_0$ and $m^\ast$ are
varied, while the maximum mass is
approximately fixed by the empirical relation previously derived for
$M_{\max}(n_0,m^\ast,\zeta)$.
As a result, the sequences belonging to one panel nearly coincide.
The coupling parameters and saturation properties of these models are
collected in Table~\ref{tab:supp-couplings} of
Appendix~\ref{app:modeldata}.

Although the nucleon-only models were calibrated to reach
$2.35\,M_\odot$, the inclusion of hyperons softens the equation of
state and reduces the maximum mass.
Rapid rotation at $707\,\mathrm{Hz}$ partially compensates for this
reduction, but all models still struggle to remain above the
observational lower bound by PSR~J0952$-$0607.
A useful systematic trend is nevertheless observed when comparing the
groups: the reduction in $M_{\max}$ becomes less pronounced for larger
values of the vector self-coupling parameter $\zeta$.

The behavior of the M1--M10 sequences therefore motivates extending the
analysis to larger nucleon-only TOV maximum masses,
$M_{N,\mathrm{TOV}}$.
The systematic dependence on $\zeta$ further suggests exploring multiple
values of $\zeta$ at each baseline.
To construct stiffer models efficiently, we make use of the empirical
relation previously derived for the nucleonic maximum
mass~\cite{Nam2026},
\begin{equation} \label{eq:emp}
    M_{N,\mathrm{TOV}}(n_0,m^\ast,\zeta)
    = m^\ast f(\zeta) + n_0\,g(\zeta) + h(\zeta),
\end{equation}
where the quadratic coefficient functions are given by
\begin{equation*}
\begin{aligned}
f(\zeta)={}&-1.23900\times10^{5}\,\zeta^2
              +1084.4\,\zeta-3.6812,\\
g(\zeta)={}&-3.12652\times10^{5}\,\zeta^2
              +2681.2\,\zeta-8.4824,\\
h(\zeta)={}&\phantom{-}1.48575\times10^{5}\,\zeta^2
              -1335.3\,\zeta+6.1959.
\end{aligned}
\end{equation*}
The functions $f$ and $h$ are in units of $M_\odot$, and $g$ is in units of
$M_\odot\,\mathrm{fm^3}$.
The relation allows us to estimate the range of the vector
self-coupling parameter $\zeta$ compatible with a given target
$M_{N,\mathrm{TOV}}$.

Figure~\ref{fig:ZetaRange} shows the resulting $\zeta$ values required
to reproduce a specified maximum mass.
Since the maximum mass increases as the saturation properties $n_0$ and
$m^\ast$ decrease, the widest possible $\zeta$ range is obtained by
adopting their lowest physically reasonable values.
Following the conditions used in our previous study where the
empirical relation was established, we therefore take $n_0 =
0.145\,\mathrm{fm^{-3}}$ and $m^\ast = 0.54$ as reference values in
estimating the allowed $\zeta$ range~\cite{Nam2026}.
As shown in Fig.~\ref{fig:ZetaRange}, a target mass of
$M_{N,\mathrm{TOV}}=2.4\,M_\odot$ allows values of $\zeta$ up to
approximately $0.002$.
However, larger target masses significantly restrict the allowed
range.
For $M_{N,\mathrm{TOV}}=2.5\,M_\odot$, values as large as
$\zeta=0.002$ are no longer compatible with the empirical relation,
and at still higher baselines even $\zeta=0.001$ approaches the upper
limit.
Guided by these considerations, we construct three sets of new RMF models.
The first set (T1--T3) corresponds to
$M_{N,\mathrm{TOV}}=2.35\,M_\odot$ and the second (T4--T6) to
$2.4\,M_\odot$, both with $\zeta=0$, $0.001$, and $0.002$.
The third set (T7--T9) corresponds to $M_{N,\mathrm{TOV}}=2.5\,M_\odot$,
for which we adopt $\zeta=0$, $0.0005$ and $0.001$.
Among the T1--T9 models, T6 has $m^\ast=0.534$, slightly below the
nominal lower value $m^\ast=0.54$.
This choice was made to keep the saturation density fixed at
$n_0=0.148\,\mathrm{fm}^{-3}$ across the T1--T9 set at the expense of
a small extension toward lower $m^\ast$.
Such an extension does not compromise the consistency of the analysis.
As demonstrated in Appendix~\ref{app:n0}, for a given value of $\zeta$,
different combinations of $n_0$ and $m^\ast$ that reproduce the same
nucleon-only TOV maximum mass, $M_{N,\mathrm{TOV}}$, lead to only minor
differences in the resulting stellar properties.
We therefore favor the uniform choice of $n_0$ across T1--T9, which
provides better control of the model comparison, despite the slightly
smaller $m^\ast$ required for T6.
The RMF parameters and corresponding nuclear-matter properties of the
T1--T9 models are summarized in Table~\ref{tab:T1-9}.

The construction of the new model sets follows the same
strategy used for the M1--M10 models.
Specifically, binding energy per particle and incompressibility are
set to $B/A=-16.3\,\mathrm{MeV}$ and
$K=240\,\mathrm{MeV}$~\cite{dutra12,Khan:2013mga}.
In particular, the EOS at low densities is required to respect the
chiral effective field theory constraints on pure neutron matter up to
$n \simeq
0.21\,\mathrm{fm^{-3}}$~\cite{hebeler13,tews13,drischler16}.
% tall 3-panel float: "!" is needed to override \topfraction
\begin{figure}[!t]
    \centering
    \includegraphics[width=1.0\linewidth]{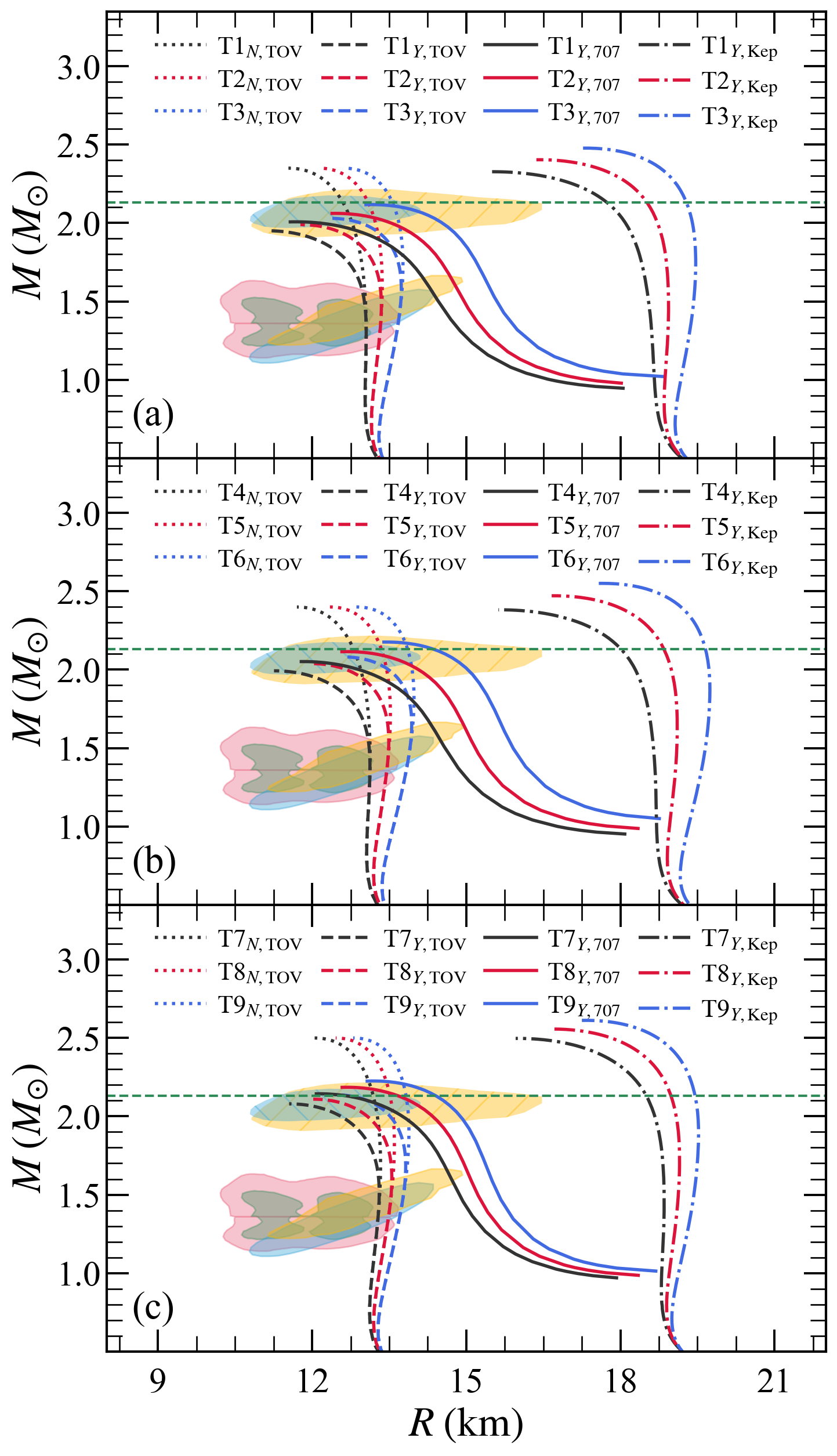}
    \caption{Mass--radius relations for T1--T9, grouped by the
    nucleonic baseline mass: (a) T1--T3 with
    $M_{N,\mathrm{TOV}}=2.35\,M_\odot$, (b) T4--T6 with
    $2.40\,M_\odot$, and (c) T7--T9 with $2.50\,M_\odot$.
    Curve styles and observational regions follow
    Fig.~\ref{fig:M1-10}; within each panel, $\zeta$ increases with the
    model number.}
    \label{fig:T1-9}
\end{figure}

\begin{table*}[t]
\caption{RMF couplings and symmetric-nuclear-matter properties at
saturation for T1--T9, grouped by the nucleonic baseline mass.
All models use $m=939$ MeV, $m_\sigma=508.194$ MeV,
$m_\omega=782.5$ MeV, and $m_\rho=763$ MeV.
The tabulated values of $\kappa$, $\lambda$, and $\zeta$ are in units of
$10^{-3}$, and those of $\Lambda_{s1}$ and $\Lambda_{s2}$ in units of
$10^{-2}$.
The saturation density $n_0$ is in fm$^{-3}$; $B/A$, $K$, $Q$, $J$,
and $L$ are in MeV and denote the binding energy per particle,
incompressibility, skewness, symmetry energy, and symmetry-energy
slope, respectively.
The dimensionless nucleon effective mass at saturation is
$m^\ast=m_N^\ast(n_0)/m$.\label{tab:T1-9}}

\begin{ruledtabular}
\scriptsize
\setlength{\tabcolsep}{3pt}
\renewcommand{\arraystretch}{1.12}

\begin{tabular}{ccccccccc|ccccccc}

 & \multicolumn{8}{c|}{Coupling parameters} & \multicolumn{7}{c}{Nuclear matter properties} \\

Model & $g_\sigma$ & $g_\omega$ & $g_\rho$
& $\kappa ~(\rm fm^{-1})$ & $\lambda$ & $\zeta$
& $\Lambda_{s1} ~(\rm fm^{-1})$ & $\Lambda_{s2}$
& $n_0$ & $m^\ast$ & $B/A$ & $K$ & $Q$ & $J$ & $L$ \\
\hline
T1 & \phantom{0}9.168 & 10.783 & 12.184 & 21.01 & $-4.44$ & 0
& \phantom{-}1.820 & 1.340 & 0.148 & 0.701 & $-16.3$ & 240 & $-$422.0 & 33.17 & 61.85 \\

T2 & \phantom{0}9.860 & 12.131 & 12.039 & 12.94 & $-2.75$ & 1
& \phantom{-}1.078 & 1.469 & 0.148 & 0.641 & $-16.3$ & 240 & $-$307.2 & 32.91 & 61.41 \\

T3 & 10.888 & 13.893 & 12.295 & \phantom{0}8.38 & $-1.52$ & 2
& \phantom{-}0.284 & 1.766 & 0.148 & 0.563 & $-16.3$ & 240 & \phantom{-}266.7 & 32.50 & 62.85 \\

T4 & \phantom{0}9.298 & 11.038 & 12.175 & 19.27 & $-4.28$ & 0
& \phantom{-}1.714 & 1.336 & 0.148 & 0.690 & $-16.3$ & 240 & $-$403.0 & 33.12 & 61.69 \\

T5 & 10.074 & 12.515 & 12.053 & 11.61 & $-2.57$ & 1
& \phantom{-}0.896 & 1.519 & 0.148 & 0.622 & $-16.3$ & 240 & $-$215.2 & 32.80 & 61.38 \\

T6 & 11.310 & 14.517 & 12.686 & \phantom{0}8.02 & $-1.52$ & 2
& \phantom{-}0.093 & 1.892 & 0.148 & 0.534 & $-16.3$ & 240 & \phantom{-}711.3 & 32.34 & 64.39 \\

T7 & \phantom{0}9.562 & 11.544 & 12.157 & 16.32 & $-3.92$ & 0
& \phantom{-}1.486 & 1.355 & 0.148 & 0.665 & $-16.3$ & 240 & $-$350.4 & 33.02 & 61.39 \\

T8 & 10.002 & 12.372 & 12.097 & 12.40 & $-3.01$ & 0.5
& \phantom{-}1.013 & 1.478 & 0.148 & 0.626 & $-16.3$ & 240 & $-$220.5 & 32.82 & 61.23 \\

T9 & 10.542 & 13.314 & 12.163 & \phantom{0}9.66 & $-2.28$ & 1
& \phantom{-}0.505 & 1.666 & 0.148 & 0.581 & $-16.3$ & 240 & \phantom{-}100.9 & 32.56 & 61.88 \\

\end{tabular}
\end{ruledtabular}
\end{table*}

Panel (a) of Fig.~\ref{fig:T1-9} shows the mass--radius relations for
T1--T3.
Across the three models, the reduction of the maximum mass induced by
hyperons becomes progressively weaker as $\zeta$ increases.
Correspondingly, $M_{Y,707}$ increases from $2.01\,M_\odot$ for T1 to
$2.06\,M_\odot$ for T2 and $2.12\,M_\odot$ for T3.
Even T3, with $\zeta=0.002$, remains slightly below the $2\sigma$ lower
bound of $2.13\,M_\odot$ inferred from PSR~J0952$-$0607.
For comparison with the GW170817~\cite{abbott2018b} and
NICER~\cite{Miller2019,Riley2019,Miller2021,Riley2021} constraints shown
in Fig.~\ref{fig:T1-9}, the nonrotating hyperonic TOV sequences are used.
This is a reasonable approximation for the NICER pulsars considered here,
since PSR~J0030+0451 and PSR~J0740+6620 rotate at approximately $205$ and
$346\,\mathrm{Hz}$, respectively, and rotational corrections to their
mass--radius relations are relatively small.
For a canonical $1.4\,M_\odot$ neutron star, $R_{1.4}$ increases from T1
to T3.
All three models are compatible with the radius constraints inferred from
the two independent NICER analyses of PSR~J0030+0451.

The GW170817 constraint provides a stronger discrimination among them: T1
lies within the $90\%$ credible region, T2 is compatible with the $50\%$
credible region, whereas the larger $R_{1.4}$ predicted by T3 is difficult
to reconcile with the GW170817 constraint.
At the high-mass end, the same hyperonic TOV sequences can be compared
with the two independent NICER analyses of PSR~J0740+6620.
T1 is compatible with the credible region inferred by Miller et
al.~\cite{Miller2021} but does not reach that obtained by Riley et
al.~\cite{Riley2021}, whereas T2 and T3 are compatible with the
constraints from both analyses.

Panel (b) of Fig.~\ref{fig:T1-9} shows the mass--radius relations for
T4--T6.
The weakening of the hyperon-induced softening with increasing $\zeta$
continues in this set, with $M_{Y,707}$ increasing from $2.05\,M_\odot$
for T4 to $2.12\,M_\odot$ for T5 and $2.18\,M_\odot$ for T6.
Consequently, T6 lies above the $2.13\,M_\odot$ lower bound from
PSR~J0952$-$0607, while T4 and T5 remain below it.
The canonical radius again increases from T4 to T6.
All three models are compatible with the radius constraints inferred
from both NICER analyses of PSR~J0030+0451.
The GW170817 constraint is more restrictive: T4 is compatible with the
$50\%$ credible region, T5 lies within the $90\%$ credible region,
whereas the larger $R_{1.4}$ of T6 falls outside the GW170817 region.
At the high-mass end, the hyperonic TOV sequences of T4--T6 are
compatible with the constraints from both NICER analyses of
PSR~J0740+6620.

Panel (c) of Fig.~\ref{fig:T1-9} shows the mass--radius relations for
T7--T9.
Unlike the preceding two sets, all three models satisfy the
PSR~J0952$-$0607 lower bound, with $M_{Y,707}=2.14$, $2.19$, and
$2.23\,M_\odot$ for T7, T8, and T9, respectively.
The same systematic reduction of hyperon-induced softening with
increasing $\zeta$ is therefore also visible in this set.
The canonical radii continue to increase from T7 to T9.
T7 and T8 are compatible with the $90\%$ credible region from GW170817,
whereas the larger $R_{1.4}$ of T9 lies outside this region.
All three models remain compatible with the radius constraints from both
NICER analyses of PSR~J0030+0451.
At the high-mass end, the hyperonic TOV sequences of T7--T9 are likewise
compatible with the credible regions inferred from both NICER analyses of
PSR~J0740+6620.

Considering the overall results of the T1--T9 parameter sets, at fixed
$M_{N,\mathrm{TOV}}$, increasing $\zeta$ leads to a larger maximum mass
supported by the hyperonic EOS, while also increasing the predicted
neutron-star radii.
Within the present model construction, the former trend facilitates
consistency with observational lower bounds on the maximum mass, whereas
the latter can be less favorable for constraints that prefer smaller
radii.
The maximum-mass and radius constraints therefore respond to variations
in $\zeta$ in opposite directions.

The comparison based on $R_{1.4}$ should, however, be interpreted with
some caution.
In addition to its dependence on $\zeta$, $R_{1.4}$ is sensitive to the
density dependence of the symmetry energy, particularly the slope
parameter $L$, which is restricted to a relatively narrow range in the
T1--T9 models.
A different choice of the isovector sector could therefore shift the
predicted canonical radii.
Moreover, $R_{1.4}$ retains some dependence on the treatment of the
low-density crust.
The radii reported here are obtained using the BPS EOS~\cite{BPS} in the
crust region, and some quantitative variation may arise with a different
crust prescription.

An additional feature can be observed when comparing the models T1,
T4, and T7.
These models correspond to the cases with $\zeta=0$, but are
constructed with different nucleon-only TOV maximum masses.
Interestingly, once hyperons are included and the stars are spun up to
the mass-shedding (Kepler) limit, the reduction of the maximum mass
caused by hyperon softening is nearly compensated by the increase due
to rapid rotation, so that $M_{N,\mathrm{TOV}} \approx
M_{Y,\mathrm{Kep}}$.

\begin{figure*}[!t]
    \centering
    \includegraphics[width=\textwidth]{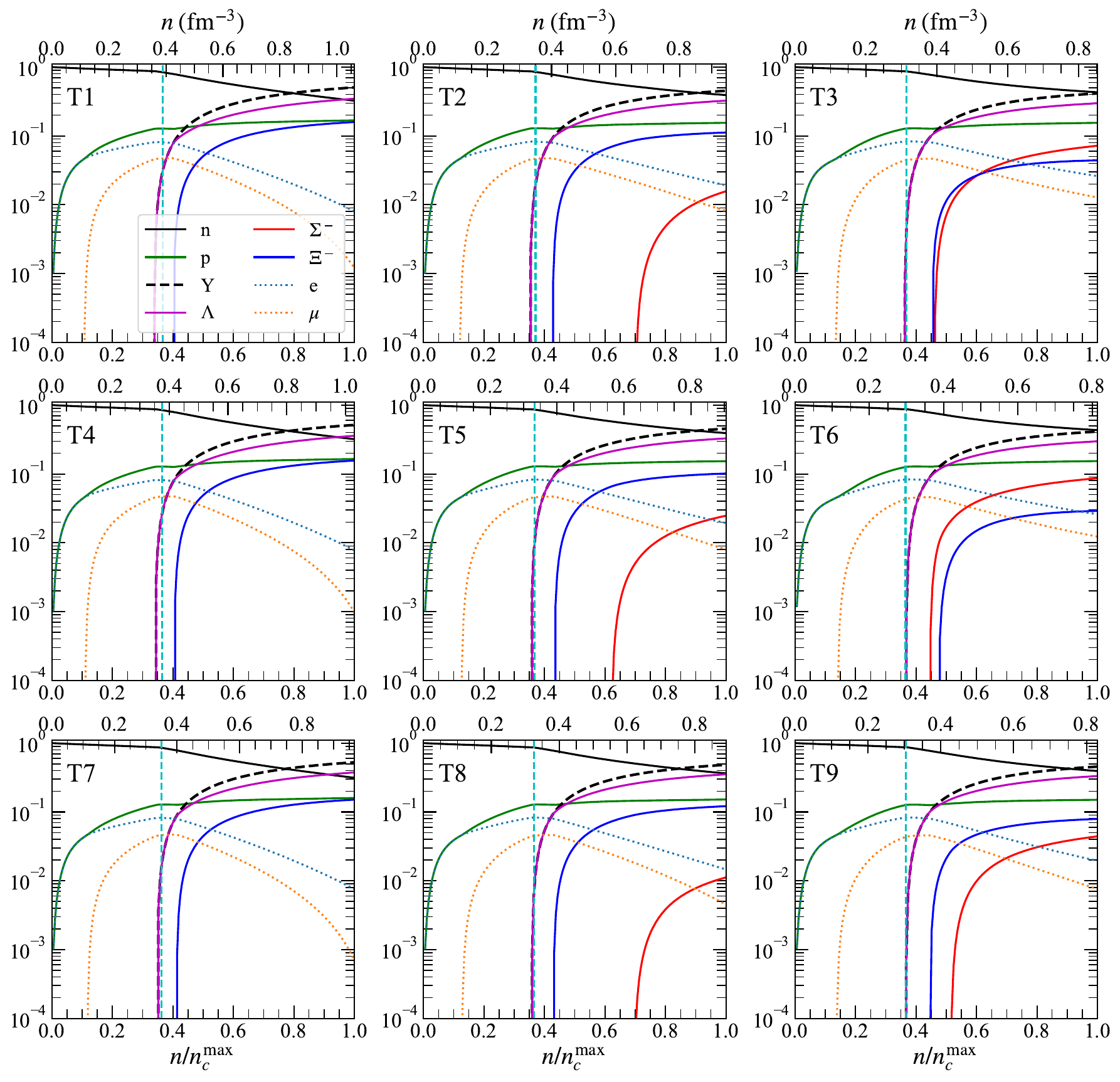}
    \caption{Particle fractions $y_i$ for the models T1--T9, plotted against 
    baryon density normalized to $n_c^{\mathrm{max}}$, the central density of 
    the nonrotating hyperonic maximum-mass configuration of the same model. 
    The upper axis gives the corresponding unnormalized density $n$ in $\mathrm{fm^{-3}}$.
    The vertical dashed line marks the central density of the
    $1.4\,M_\odot$ configuration for the same model.}
    \label{fig:Fraction_T1-9}
\end{figure*}

\begin{table}[!b]
\caption{Maximum masses and central hyperon fractions for T1--T9.
$M_{N,\mathrm{TOV}}$ is the nucleonic nonrotating maximum mass, while
$M_{Y,\mathrm{TOV}}$, $M_{Y,707}$, and $M_{Y,\mathrm{Kep}}$ are the
hyperonic maximum masses for nonrotating, $707\,\mathrm{Hz}$, and
Kepler-limit configurations, respectively.
$y_Y(n_c^{\mathrm{max}})$ is the total hyperon fraction at the center
of the configuration defining $M_{Y,\mathrm{TOV}}$.
Masses are in $M_\odot$.}
\begin{ruledtabular}
\begin{tabular}{cccccc}
Model & $M_{N,\mathrm{TOV}}$ & $M_{Y,\mathrm{TOV}}$ & $M_{Y,707}$ & $M_{Y,\mathrm{Kep}}$ & $y_Y(n_c^{\mathrm{max}})$ \\
\hline
T1 & 2.35 & 1.95 & 2.01 & 2.33 & 0.507 \\
T2 & 2.35 & 1.99 & 2.06 & 2.40 & 0.453 \\
T3 & 2.35 & 2.03 & 2.12 & 2.48 & 0.415 \\
T4 & 2.40 & 1.99 & 2.05 & 2.38 & 0.515 \\
T5 & 2.40 & 2.04 & 2.12 & 2.47 & 0.455 \\
T6 & 2.40 & 2.08 & 2.18 & 2.55 & 0.414 \\
T7 & 2.50 & 2.08 & 2.14 & 2.50 & 0.527 \\
T8 & 2.50 & 2.11 & 2.19 & 2.56 & 0.485 \\
T9 & 2.50 & 2.14 & 2.23 & 2.61 & 0.456 \\

\end{tabular}
\end{ruledtabular}
\label{tab:T1-9Mmax}
\end{table}

To understand the different degrees of hyperon softening among the
models, the particle fractions are examined as functions of baryon
density in Fig.~\ref{fig:Fraction_T1-9}.
The density range extends to the central density of the corresponding
nonrotating hyperonic maximum-mass configuration.
The associated maximum masses and total hyperon fractions at these
central densities are summarized in Table~\ref{tab:T1-9Mmax}.

The values in Table~\ref{tab:T1-9Mmax} provide a direct numerical
comparison of the central compositions.
In order of increasing $\zeta$, the central hyperon fraction decreases
from $0.507$ through $0.453$ to $0.415$ across T1--T3 and from $0.515$
through $0.455$ to $0.414$ across T4--T6.
The same behavior is found across T7--T9, for which the fraction
decreases from $0.527$ through $0.485$ to $0.456$.
These values show that the hyperon population near the stellar center
becomes systematically smaller as $\zeta$ increases, thereby weakening
the hyperon-induced softening of the EOS.

The vertical dashed lines in Fig.~\ref{fig:Fraction_T1-9} mark the
central densities of the corresponding $1.4\,M_\odot$ configurations.
Below these densities, hyperons are either absent or only weakly
populated, indicating that canonical-mass stars remain essentially
nucleonic in all nine models.
This explains why the hyperonic mass--radius sequences in
Fig.~\ref{fig:T1-9} closely follow their nucleon-only counterparts up
to the canonical-mass region.

The ordering of the hyperon thresholds provides an
additional distinction among the models.
In T5, the $\Xi^{-}$ hyperon appears before the $\Sigma^{-}$, while in
T4 the $\Sigma^{-}$ does not reach its threshold at all below the
central density of the maximum-mass configuration, so that only the
$\Xi^{-}$ is populated.
As $\zeta$ increases, the onset density of the $\Sigma^{-}$ approaches
that of the $\Xi^{-}$, and their ordering is reversed in T6. 
For the model sets with larger $M_{N,\rm TOV}$,
%\tcm{(JH: I think ``higher-baseline sets'' is ambiguous. Maybe instead: ``For the model sets with larger nucleon-only TOV maximum masses"?} \tcb{GH: Yes. I agree. I switched it}
the explored $\zeta$ range is smaller, and a full inversion of the onset order does
not occur.
Nevertheless, the density gap between the $\Xi^{-}$ and $\Sigma^{-}$
onsets is visibly reduced as $\zeta$ increases, indicating the same
underlying trend.

\begin{figure*}[!t]
    \centering
    \includegraphics[width=\textwidth]{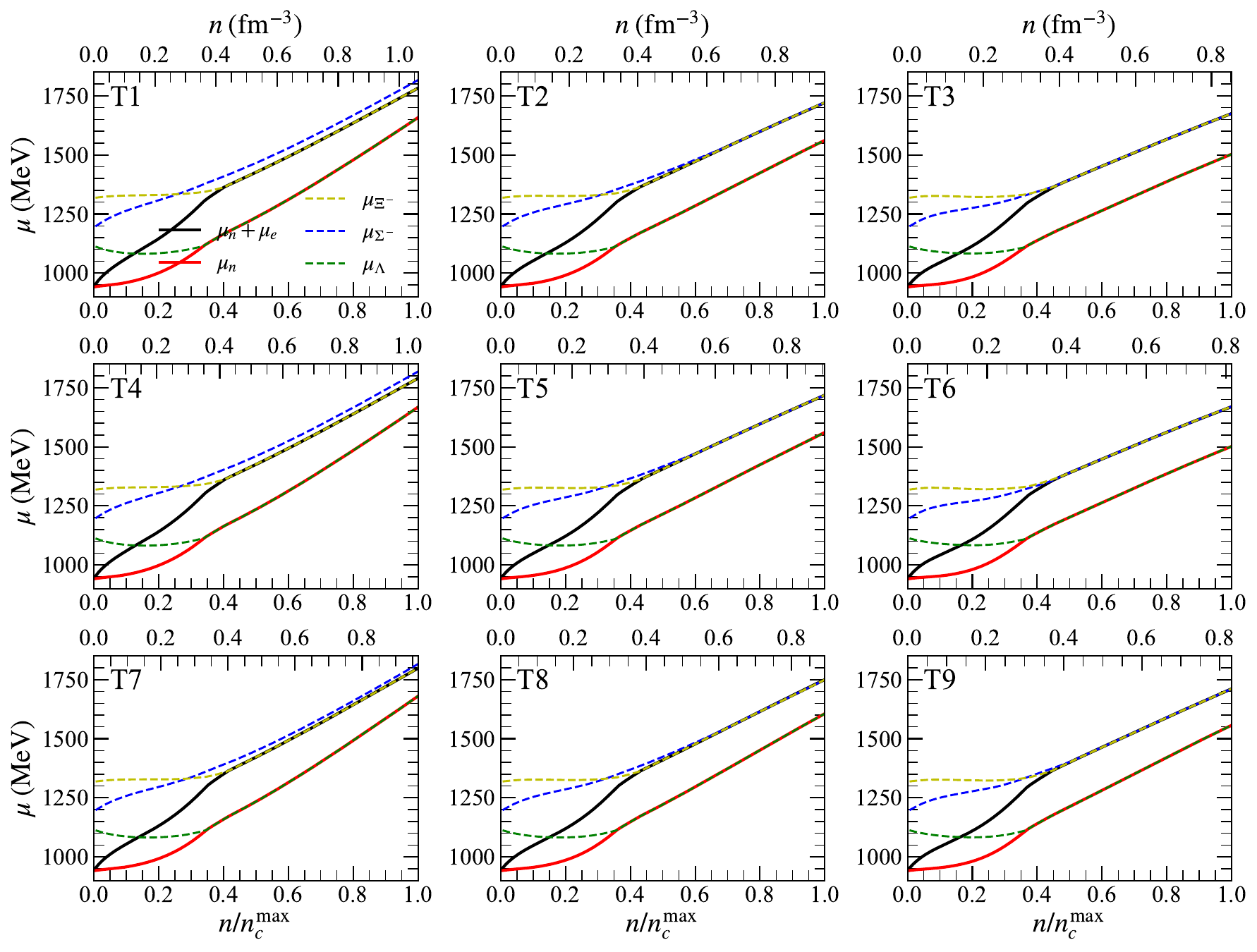}
    \caption{Chemical potentials $\mu$ for the models T1--T9, plotted against 
    baryon density normalized to $n_c^{\mathrm{max}}$ as in 
    Fig.~\ref{fig:Fraction_T1-9}; the upper axis gives the corresponding 
    unnormalized density $n$ in $\mathrm{fm^{-3}}$.
    Solid curves show the reference potentials $\mu_n+\mu_e$ and $\mu_n$,
    and dashed curves show the hyperon potentials $\mu_\Lambda$,
    $\mu_{\Sigma^-}$ and $\mu_{\Xi^-}$.
    }

    \label{fig:ChemPot_T1-9}
\end{figure*}

Figure~\ref{fig:ChemPot_T1-9} shows the density dependence of the
relevant chemical potentials corresponding to
Fig.~\ref{fig:Fraction_T1-9}.
The onset condition for hyperons is determined by
Eq.~\eqref{eq:beta_eq}.
Therefore, the relative positions of $\mu_{\Sigma^-}$, $\mu_{\Xi^-}$,
and the reference curve $\mu_n+\mu_e$ directly explain the hyperon
onset sequence.
In the T4 model, $\mu_{\Sigma^-}$ increases rapidly with density and
exceeds $\mu_{\Xi^-}$ even before the $\Lambda$ onset.
However, $\mu_{\Sigma^-}$ does not reach the threshold condition
$\mu_n+\mu_e$ up to the central density of the maximum-mass
configuration.
Thus, $\Sigma^-$ does not appear in the stable neutron-star branch for
T4.
In T5, a similar inversion between $\mu_{\Sigma^-}$ and $\mu_{\Xi^-}$
occurs, but it takes place at a higher density than in T4.
After the inversion, the two chemical potentials approach each other
rather quickly, resulting in a less pronounced separation between the
two thresholds.
In contrast, in T6, such an inversion does not occur within the
relevant density range, and $\mu_{\Sigma^-}$ reaches the threshold
condition earlier than $\mu_{\Xi^-}$.
This leads to the earlier appearance of $\Sigma^-$.
This behavior can be understood from the RMF expression for the
hyperon chemical potential at the onset,
\begin{equation}
\mu_Y(k=0)
=m_Y^\ast+ g_{\omega Y}\omega_0+ g_{\phi Y}\phi_0
+ g_\rho I_{3Y}\rho_{03}.
\end{equation}
Before the appearance of hyperons, the $\phi$-meson contribution
vanishes, since the $\phi$ field is generated only by hyperons.
Hence, the dominant density-dependent vector contribution is given by
$g_{\omega Y}\omega_0$.
The quartic $\omega$-meson self-interaction, controlled by $\zeta$,
suppresses the growth of $\omega_0$.
This reduction affects $\mu_{\Sigma^-}$ more strongly than
$\mu_{\Xi^-}$ because the SU(6) relation gives
$g_{\omega\Sigma}=2g_{\omega\Xi}$.
Consequently, the relative ordering of $\mu_{\Sigma^-}$ and
$\mu_{\Xi^-}$ is sensitive to the strength of the $\omega$-meson
self-interaction.

\begin{figure*}[!t]
    \centering
    \includegraphics[width=\textwidth]{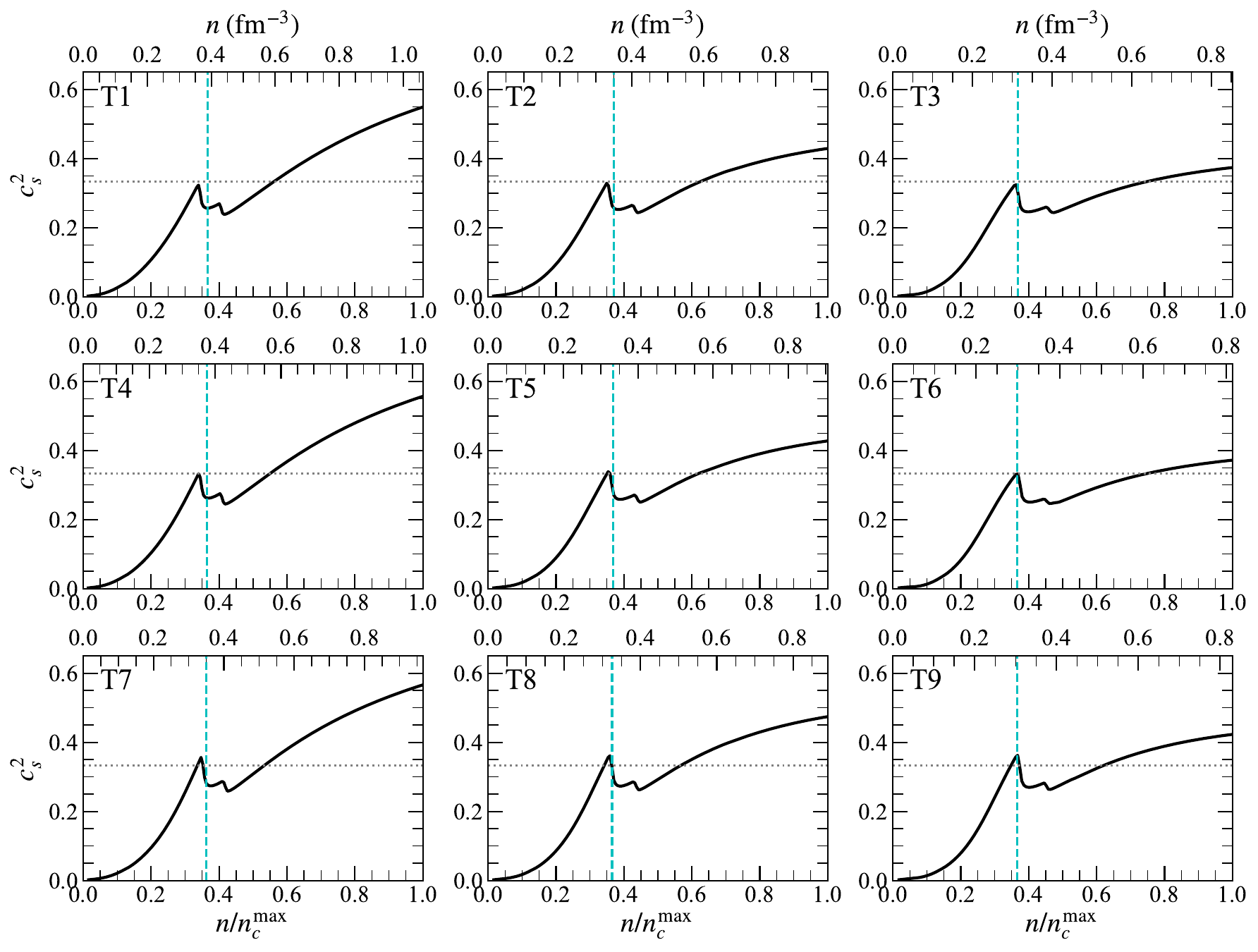}
    \caption{Squared speed of sound $c_s^2$, defined in
    Squared speed of sound $c_s^2$, defined in 
    Eq.~\eqref{eq:sound_speed}, for T1--T9, plotted against baryon density 
    normalized to $n_c^{\mathrm{max}}$ as in Fig.~\ref{fig:Fraction_T1-9}; 
    the upper axis gives the corresponding unnormalized density $n$ in $\mathrm{fm^{-3}}$.
    The vertical dashed lines mark the central densities of the
    $1.4\,M_\odot$ configurations.
    The horizontal dotted line marks the conformal value $c_s^2=1/3$.}
    \label{fig:soundvel}
\end{figure*}

At fixed $M_{N,\mathrm{TOV}}$ and $n_0$, $\zeta$ and $m^\ast$ cannot be
varied independently: through the empirical relation~\eqref{eq:emp}, a
larger $\zeta$ requires a smaller $m^\ast$.
The smaller $m^\ast$ corresponds to a stronger scalar mean field, which
reduces the hyperon Dirac effective mass,
$m_Y^\ast=m_Y-g_{\sigma Y}\sigma$, and
thereby favors an earlier onset.
This effect is particularly important for the competition between
$\Sigma^-$ and $\Xi^-$.
At $\zeta=0$, the repulsive $\Sigma$ potential makes $\Sigma^-$ less
favorable than $\Xi^-$ despite its smaller bare mass.
As $\zeta$ increases, however, the stronger scalar field lowers
$m_\Sigma^*$ more rapidly and shifts the $\Sigma^-$ onset toward
substantially lower density.
Because $\Sigma^-$ is lighter than $\Xi^-$, this shift can overcome the
initial disadvantage associated with its repulsive potential.
The resulting change should therefore be understood as a rapid advance
of the $\Sigma^-$ onset, rather than as $\Xi^-$ becoming less
favorable.

The same composition changes are imprinted on the speed of sound,
\begin{equation}
c_s^2=\frac{dP}{d\varepsilon},
\label{eq:sound_speed}
\end{equation}
which is shown for the hyperonic T1--T9 EOSs in
Fig.~\ref{fig:soundvel}.
Below the first hyperon threshold the matter is nucleonic and $c_s^2$
rises steeply, reaching a local maximum of $0.32$--$0.36$ at
$n\simeq0.30$--$0.36\,\mathrm{fm^{-3}}$.
Each hyperon onset then converts part of the pressure-carrying Fermi
sea into a new, initially nonrelativistic species and produces a local
minimum in $c_s^2$. The first accompanies the $\Lambda$ threshold; the
second is set by whichever of $\Xi^-$ and $\Sigma^-$ appears next, so it
follows the onset ordering discussed above.
In T3 the two onsets lie within $0.005\,\mathrm{fm^{-3}}$ of each other, and
in T6 the ordering is reversed with $\Sigma^-$ first; in both the two
contributions to the second minimum cannot be separated.
The resulting dip, $c_s^2\simeq0.24$--$0.27$ over $0.32\lesssim
n\lesssim0.46\,\mathrm{fm^{-3}}$, is the softening that lowers
$M_{Y,\mathrm{TOV}}$ relative to $M_{N,\mathrm{TOV}}$.
Once the hyperon populations saturate, the vector repulsion again
dominates and $c_s^2$ increases monotonically, crossing the conformal
value $1/3$ between $n\simeq0.51$ and
$0.64\,\mathrm{fm^{-3}}$~\cite{bedaque15,Annala:2020puf}.

\begin{figure*}[t]
    \centering
    \includegraphics[width=\textwidth]{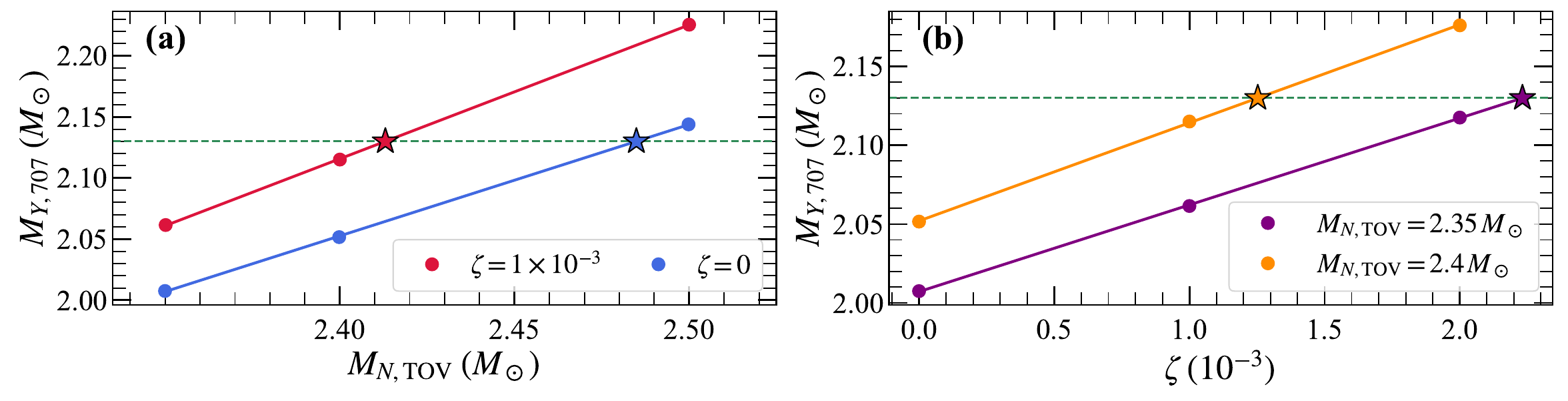}
    \caption{Threshold pairs obtained from the lower mass threshold
    $M_{Y,707}=2.13\,M_\odot$.
    (a) $M_{Y,707}$ as a function of $M_{N,\mathrm{TOV}}$ at fixed
    $\zeta$ and (b) as a function of $\zeta$ at fixed
    $M_{N,\mathrm{TOV}}$.
    Circles show the T1--T9 results, solid lines
    are linear interpolations, horizontal dashed lines mark the
    threshold, and stars mark the intersections.}
    \label{fig:NMrange-intersections}
\end{figure*}

We now examine the relation between $M_{Y,707}$ and the nucleonic
TOV maximum mass $M_{N,\mathrm{TOV}}$ for fixed values of $\zeta$,
using the maximum masses of Table~\ref{tab:T1-9Mmax} to constrain the
nuclear-matter properties.
As shown in Fig.~\ref{fig:NMrange-intersections}(a), for each fixed $\zeta$ the
values of $M_{Y,707}$ vary approximately linearly with
$M_{N,\mathrm{TOV}}$.
By fitting these relations linearly and interpolating to the condition
$M_{Y,707}=2.13\,M_\odot$, we extract the threshold values of
$M_{N,\mathrm{TOV}}$: $(\zeta,\,M_{N,\mathrm{TOV}}) =
(0,\,2.485),\quad (10^{-3},\,2.413).$ Similarly, as shown in
Fig.~\ref{fig:NMrange-intersections}(b), $M_{Y,707}$ varies approximately linearly
with $\zeta$ for fixed $M_{N,\mathrm{TOV}}$.
Applying a linear fit to the computed data points and intersecting
with $M_{Y,707}=2.13\,M_\odot$, we obtain the threshold values of
$\zeta$: $(\zeta,\,M_{N,\mathrm{TOV}}) =
(1.25\times10^{-3},\,2.40),\quad (2.23\times10^{-3},\,2.35)$.
Fitting the four threshold pairs obtained from the two panels with a
quadratic polynomial yields the boundary curve
\begin{equation}
    M_{N,\mathrm{TOV}}(\zeta) = 8.38\times10^{3}\,\zeta^2 - 79.02\,\zeta + 2.485.
    \label{eq:boundary}
\end{equation}

\begin{figure*}[t]
    \centering
    \includegraphics[width=\textwidth]{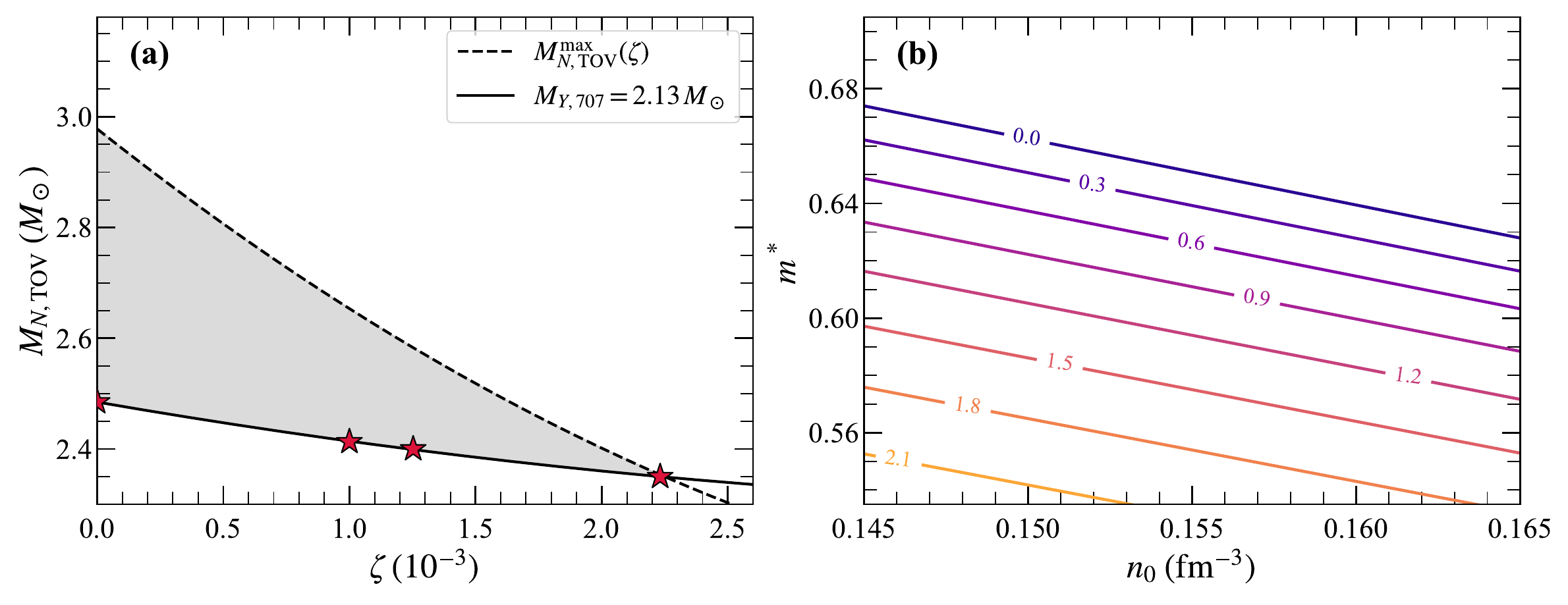}
    \caption{Constraints implied by the boundary obtained in
    Fig.~\ref{fig:NMrange-intersections}.
    (a) Allowed region (shaded) between the minimum
    $M_{N,\mathrm{TOV}}$ required by the mass threshold (solid) and the
    maximum attainable $M_{N,\mathrm{TOV}}^{\max}(\zeta)$ (dashed);
    crimson stars are the threshold pairs.
    (b) Corresponding upper boundaries on the dimensionless nucleon
    effective mass at saturation, $m^\ast_{ub}(n_0,\zeta)$; curve labels
    give $\zeta$ in units of $10^{-3}$.}
    \label{fig:NMrange-constraints}
\end{figure*}

Next, we employ the empirical RMF relation for the maximum mass,
Eq.~\eqref{eq:emp}.
As shown in Fig.~\ref{fig:NMrange-constraints}(a), the boundary curve
$M_{N,\mathrm{TOV}}(\zeta)$ and the maximum mass curve
$M_{N,\mathrm{TOV}}^{\max}(\zeta)$ are plotted together in the
$(\zeta, M_{N,\mathrm{TOV}})$ plane, where the latter represents the
maximum achievable mass at each $\zeta$ obtained by fixing $n_0 =
0.145\,\mathrm{fm}^{-3}$ and $m^\ast = 0.54$, following the same
procedure as in Fig.~\ref{fig:ZetaRange}.
The shaded region between the two curves represents the EOSs satisfying
$M_{Y,707} \ge 2.13\,M_\odot$.
The two curves intersect at $(\zeta,\,M_{N,\mathrm{TOV}}) \approx
(2.25\times10^{-3},\;2.349\,M_\odot),$ beyond which the boundary curve
lies above the maximum mass curve: for $\zeta \gtrsim
2.25\times10^{-3}$ no EOS can both reach
$M_{N,\mathrm{TOV}}^{\max} \ge M_{N,\mathrm{TOV}}(\zeta)$ and satisfy
$M_{Y,707} \ge 2.13\,M_\odot$, and the allowed region closes.

As a consistency check, we explicitly constructed RMF EOSs for three
of the threshold pairs, $(\zeta,M_{N,\mathrm{TOV}})=(0,2.485)$,
$(10^{-3},2.413)$, and $(1.25\times10^{-3},2.40)$, and computed the
corresponding hyperonic configurations rotating at
$707\,\mathrm{Hz}$, shown in Fig.~\ref{fig:Ymodels}. The resulting values of $M_{Y,707}$, listed
in Table~\ref{tab:Ymass}, were found to reproduce the imposed
threshold $2.13\,M_\odot$ to within $0.002\,M_\odot$, confirming that
the threshold points inferred from the interpolation procedure are
realized by explicit EOS constructions.
The remaining point, $(2.23\times10^{-3},2.35)$, was not checked because it
lies too close to the closing point of the allowed region,
$\zeta\simeq2.25\times10^{-3}$.

Substituting the boundary values $M_{N,\mathrm{TOV}}(\zeta)$ into the
empirical relation Eq.\ \eqref{eq:emp}, we derive for each fixed $\zeta$ a
linear upper boundary in the $(n_0, m^\ast)$ plane,
\begin{equation}
    m^\ast_{\rm ub}(n_0,\zeta) = a(\zeta)\,n_0 + b(\zeta),
    \label{eq:mub}
\end{equation}
with
\begin{equation}
    a(\zeta) = -\frac{g(\zeta)}{f(\zeta)},\qquad
    b(\zeta) = \frac{M_{N,\mathrm{TOV}}(\zeta)-h(\zeta)}{f(\zeta)},
    \label{eq:mub_ab}
\end{equation}
where $M_{N,\mathrm{TOV}}(\zeta)$ is the boundary curve of
Eq.~\eqref{eq:boundary}, and the coefficient functions $f(\zeta)$,
$g(\zeta)$, and $h(\zeta)$ are those entering Eq.~\eqref{eq:emp}.
As shown in Fig.~\ref{fig:NMrange-constraints}(b), each contour line represents
this upper boundary for a given value of $\zeta$.
For fixed $(n_0, \zeta)$, an RMF model is compatible with the
observational constraint if $m^\ast \le m^\ast_{\rm ub}(n_0,\zeta)$, that is,
only combinations lying below the corresponding contour line
in Fig.~\ref{fig:NMrange-constraints}(b) can support a hyperonic neutron star with
$M_{Y,707} \ge 2.13\,M_\odot$.
For example, in the case of the NL3 \cite{Lalazissis1997} model,
$\zeta=0$ and $(n_0,m^\ast)=(0.148,0.6)$, which satisfies the
condition $m^\ast < m^\ast_{\mathrm{ub}}(n_0=0.148,\zeta=0)=0.667$.
%\tcm{YH: you have to unify the expression $m^\ast_{\mathrm{ub}}$. Sometimes, you just write $m^\ast_{\rm ub}$ without `mathrm`. Sine `ub' stands for upper bound. I prefer to use `mathrm'} 
%\tcb{GH: It has been changed accordingly.}
Therefore, NL3 can be regarded as an acceptable parameter set if only
the maximum-mass constraint is considered.
In contrast, the IUFSU \cite{Fattoyev2010} and FSU
\cite{Todd-Rutel2005} models already employ $\zeta=0.005$ and $0.01$,
respectively, both above the value $0.00225$ at which the allowed region
closes, so neither satisfies the corresponding condition.

Two remarks on the scope of this construction are in order, since it rests on
a small number of equations of state.
First, every model entering Fig.~\ref{fig:NMrange-intersections} is built at the same
saturation density.  Appendix~\ref{app:n0} tests this choice with
fixed-$M_{N,\mathrm{TOV}}$ control sequences spanning
$n_0=0.148$--$0.165\,\mathrm{fm^{-3}}$ wherever the adopted lower limit on
$m^\ast$ permits; $M_{Y,707}$ changes by at most $0.017\,M_\odot$.
Second, the sequences in Fig.~\ref{fig:NMrange-intersections}(a), constructed at fixed
$\zeta$ using the three nucleonic baselines of T1--T9, and those in
Fig.~\ref{fig:NMrange-intersections}(b), constructed at fixed $M_{N,\mathrm{TOV}}$
using the corresponding $\zeta$ values, both rely on the T1--T9 models.
The higher-baseline T10--T15 models in
Appendix~\ref{app:sigma1} extend both trends and provide the bracketing
sequences needed for the $1\sigma$ sensitivity test; their masses are listed
in Table~\ref{tab:T1015mass}.

All the analyses in this section are based on the $2\sigma$ lower bound
inferred from PSR~J0952$-$0607, $M_{Y,707}\ge2.13\,M_\odot$.
The corresponding calculation at the $1\sigma$ edge of the same
measurement, $M_{Y,707}\ge2.24\,M_\odot$, is given in
Appendix~\ref{app:sigma1}.

\section{Conclusion}\label{sec:conclusion}

We have shown how the mass of a rapidly rotating pulsar can be
translated into constraints on RMF models with hyperonic cores.  The
improved mass determination of PSR~J0952$-$0607 provided the
observational impetus for this analysis.  To assess its implications,
we extended nucleonic RMF models using a fixed hyperon-coupling
prescription and computed stellar sequences at the observed spin
frequency of $707\,\mathrm{Hz}$.  This construction treats consistently
the competing effects relevant to the maximum mass: softening of the
EOS by hyperons and rotational support.

The model sequences reveal a systematic relation between the correlated
$(\zeta,m^\ast)$ sequences and the hyperon content.  At fixed
$M_{N,\mathrm{TOV}}$, increasing $\zeta$ together with the correlated
decrease of $m^\ast$ reduces the central hyperon fraction and therefore
weakens the reduction of the maximum mass.  The same correlated change
shifts the relative thresholds of $\Sigma^-$ and $\Xi^-$ and can reverse
their onset order.  These trends cannot be attributed to $\zeta$ alone,
because $\zeta$ and $m^\ast$ are linked by the empirical maximum-mass
relation used to construct the models.  Moreover, the accompanying
increase in the canonical radius within these sequences illustrates
that satisfying a large maximum mass need not improve agreement with
radius constraints.

Our main quantitative result follows from the intersections of the
calculated T1--T9 sequences with the observational threshold.  At fixed
$\zeta$, interpolating $M_{Y,707}$ as a function of
$M_{N,\mathrm{TOV}}$ determines the minimum nucleonic baseline required
for the rotating hyperonic sequence to reach $2.13\,M_\odot$;
complementary interpolations in $\zeta$ at fixed
$M_{N,\mathrm{TOV}}$ provide two additional threshold pairs.  A
quadratic fit through the resulting four intersections defines the
boundary of Eq.~\eqref{eq:boundary},
$M_{N,\mathrm{TOV}}(\zeta)=8.38\times10^{3}\,\zeta^2
-79.02\,\zeta+2.485$.  Thus, for each $\zeta$, this curve gives the
minimum $M_{N,\mathrm{TOV}}$ needed to support a hyperonic star with
$M_{Y,707}\geq2.13\,M_\odot$.  Combined with the RMF empirical maximum-mass relation,
it yields the upper limit $m^\ast\leq m^\ast_{ub}(n_0,\zeta)$ shown in
Fig.~\ref{fig:NMrange-constraints}(b), and the allowed region closes at
$\zeta\simeq2.25\times10^{-3}$ and
$M_{N,\mathrm{TOV}}\simeq2.349\,M_\odot$.  To test whether the
intersection-derived boundary corresponds to realizable RMF models, we
constructed dedicated EOSs at three of the four threshold pairs.  Their
rotating hyperonic maximum masses reproduce the target value within
$0.002\,M_\odot$, demonstrating that the interpolated boundary is
realized by explicit EOS constructions at the tested points.

The constraints inferred here depend on the adopted SU(6) vector
couplings and hyperon potential depths.  Given
the systematic uncertainties associated with optical modeling of the
irradiated companion, we use the conservative $2\sigma$ lower mass bound
of $2.13\,M_\odot$ for PSR~J0952$-$0607 as the fiducial input.
Accordingly, these constraints should not be interpreted as
model-independent exclusions.  The framework can be updated directly
as pulsar masses and hypernuclear interactions become better
determined, providing a systematic link between observations of massive
rotating neutron stars and nuclear-matter constraints within hyperonic
RMF models.

\acknowledgments
This work were supported by the National Research Foundation
of Korea (NRF) grant funded by the Korea government (MSIT)
(No.~RS-2024-00457037) and by the Global--Learning \& Academic Research
Institution for Master's and PhD students, and Postdocs (LAMP) Program of
the National Research Foundation of Korea (NRF) grant funded by the
Ministry of Education (No.~RS-2024-00442483).
P. Thakur was also supported (in part) by the Yonsei University Research Fund(Yonsei University Frontier Fellowship for
Postdoctoral Researchers) of 2025. 
J.W.H. was supported in part by the National Science Foundation under
Grant No.~PHY-2514930.

    \bibliographystyle{apsrev4-2}
    \bibliography{ref}
\appendix
% appendix floats get their own per-appendix sequence (A1, B1, ...)
\numberwithin{figure}{section}
\numberwithin{table}{section}
\renewcommand{\thefigure}{\thesection\arabic{figure}}
\renewcommand{\thetable}{\thesection\arabic{table}}
% float pages in the appendix are set flush to the top with a generous gap
% between floats, so the collected tables read as a list rather than being
% vertically centred with large blanks above and below
\makeatletter
\setlength{\@fptop}{0pt}
\setlength{\@fpsep}{28pt}
\setlength{\@fpbot}{0pt plus 1fil}
% starred floats use the \@dblfp... series
\setlength{\@dblfptop}{0pt}
\setlength{\@dblfpsep}{28pt}
\setlength{\@dblfpbot}{0pt plus 1fil}
\makeatother
% placeins[section] drops a \FloatBarrier before every \section, which keeps
% each appendix's floats inside it.  The barriers are left active here so that
% no table or figure drifts past its own appendix; APS asks that page-wide
% floats keep the starred environments rather than switching the column grid,
% so the layout is handled through float placement alone.

\clearpage
\onecolumngrid
\section{EOS parameters and nuclear-matter properties}
\label{app:modeldata}

Table~\ref{tab:supp-couplings} collects the RMF couplings and the
symmetric-nuclear-matter properties at saturation for the model sets used
outside Table~\ref{tab:T1-9}: M1--M10 of Sec.~\ref{sec:results}, Y1--Y3 of
Appendix~\ref{app:Yvalidation}, and T10--T15 of Appendix~\ref{app:sigma1}.

\setlength{\LTcapwidth}{\textwidth}
\begingroup
\footnotesize
\setlength{\tabcolsep}{3pt}
\renewcommand{\arraystretch}{1.25}
\begin{longtable}{ccccccccc|ccccccc}
\caption{RMF parameters and symmetric-nuclear-matter properties
at saturation for M1--M10, the saturation-density controls T1a--T9c of
Appendix~\ref{app:n0}, Y1--Y3, and T10--T15.
All models use $m=939$ MeV, $m_\sigma=508.194$ MeV,
$m_\omega=782.5$ MeV, and $m_\rho=763$ MeV.
Notation, scaling, and units follow Table~\ref{tab:T1-9}.
\label{tab:supp-couplings}}\\
\hline\hline
 & \multicolumn{8}{c|}{Coupling parameters} & \multicolumn{7}{c}{Nuclear matter properties} \\
Model & $g_\sigma$ & $g_\omega$ & $g_\rho$
& $\kappa ~(\rm fm^{-1})$ & $\lambda$ & $\zeta$
& $\Lambda_{s1} ~(\rm fm^{-1})$ & $\Lambda_{s2}$
& $n_0$ & $m^\ast$ & $B/A$ & $K$ & $Q$ & $J$ & $L$ \\
\hline
\endfirsthead
\multicolumn{16}{l}{\textit{Table~\thetable{} (continued)}}\\
\hline\hline
 & \multicolumn{8}{c|}{Coupling parameters} & \multicolumn{7}{c}{Nuclear matter properties} \\
Model & $g_\sigma$ & $g_\omega$ & $g_\rho$
& $\kappa ~(\rm fm^{-1})$ & $\lambda$ & $\zeta$
& $\Lambda_{s1} ~(\rm fm^{-1})$ & $\Lambda_{s2}$
& $n_0$ & $m^\ast$ & $B/A$ & $K$ & $Q$ & $J$ & $L$ \\
\hline
\endhead
\hline
\multicolumn{16}{r}{\textit{continued on the next page}}\\
\endfoot
\hline\hline
\endlastfoot
M1 & 9.137 & 10.747 & 12.123 & 21.08 & $-4.50$ & 0
& 1.897 & 1.268 & 0.1496 & 0.700 & $-16.3$ & 240 & $-420.6$ & 33.40 & 63.05 \\

M2 & 9.124 & 10.784 & 11.956 & 20.32 & $-4.56$ & 0
& 2.038 & 1.086 & 0.1540 & 0.690 & $-16.3$ & 240 & $-406.8$ & 34.02 & 66.24 \\

M3 & 9.110 & 10.818 & 11.791 & 19.63 & $-4.60$ & 0
& 2.157 & 0.927 & 0.1583 & 0.680 & $-16.3$ & 240 & $-390.2$ & 34.65 & 69.51 \\
M4 & 9.750 & 12.044 & 11.735 & 12.74 & $-2.97$ & 1
& 1.559 & 1.102 & 0.1590 & 0.620 & $-16.3$ & 240 & $-201.8$ & 34.53 & 69.94 \\

M5 & 9.775 & 12.046 & 11.852 & 12.92 & $-2.90$ & 1
& 1.389 & 1.236 & 0.1546 & 0.630 & $-16.3$ & 240 & $-256.8$ & 33.88 & 66.46 \\

M6 & 9.801 & 12.048 & 11.976 & 13.14 & $-2.83$ & 1
& 1.208 & 1.384 & 0.1501 & 0.640 & $-16.3$ & 240 & $-303.4$ & 33.23 & 63.05 \\

M7 & 9.830 & 12.050 & 12.106 & 13.39 & $-2.76$ & 1
& 1.013 & 1.547 & 0.1457 & 0.650 & $-16.3$ & 240 & $-342.7$ & 32.60 & 59.69 \\
M8 & 10.951 & 13.989 & 12.337 & 8.28 & $-1.51$ & 2
& 0.228 & 1.796 & 0.1476 & 0.560 & $-16.3$ & 240 & 310.1 & 32.41 & 62.71 \\

M9 & 10.918 & 13.940 & 12.385 & 8.53 & $-1.65$ & 2
& 0.536 & 1.655 & 0.1522 & 0.550 & $-16.3$ & 240 & 463.7 & 33.11 & 66.35 \\

M10 & 10.889 & 13.894 & 12.450 & 8.82 & $-1.80$ & 2
& 0.828 & 1.527 & 0.1568 & 0.540 & $-16.3$ & 240 & 642.2 & 33.81 & 69.99 \\

\hline
T1a & \phantom{0}9.152 & 10.783 & 12.107 & 20.78 & $-4.49$ & 0
& \phantom{-}1.898 & 1.250 & 0.150 & 0.698 & $-16.3$ & 240 & $-$417.2 & 33.45 & 63.31 \\

T1b & \phantom{0}9.116 & 10.785 & 11.916 & 20.20 & $-4.58$ & 0
& \phantom{-}2.071 & 1.045 & 0.155 & 0.688 & $-16.3$ & 240 & $-$403.8 & 34.17 & 67.03 \\

T1c & \phantom{0}9.080 & 10.783 & 11.726 & 19.69 & $-4.65$ & 0
& \phantom{-}2.219 & 0.866 & 0.160 & 0.678 & $-16.3$ & 240 & $-$388.0 & 34.90 & 70.84 \\

T1d & \phantom{0}9.046 & 10.780 & 11.539 & 19.21 & $-4.71$ & 0
& \phantom{-}2.344 & 0.708 & 0.165 & 0.669 & $-16.3$ & 240 & $-$369.1 & 35.65 & 74.74 \\

T2a & \phantom{0}9.847 & 12.129 & 11.982 & 12.84 & $-2.79$ & 1
& \phantom{-}1.165 & 1.398 & 0.150 & 0.636 & $-16.3$ & 240 & $-$287.5 & 33.19 & 62.93 \\

T2b & \phantom{0}9.818 & 12.127 & 11.848 & 12.61 & $-2.86$ & 1
& \phantom{-}1.371 & 1.235 & 0.155 & 0.625 & $-16.3$ & 240 & $-$230.3 & 33.92 & 66.78 \\

T2c & \phantom{0}9.791 & 12.125 & 11.724 & 12.43 & $-2.94$ & 1
& \phantom{-}1.564 & 1.089 & 0.160 & 0.614 & $-16.3$ & 240 & $-$161.6 & 34.66 & 70.71 \\

T2d & \phantom{0}9.768 & 12.123 & 11.611 & 12.29 & $-3.02$ & 1
& \phantom{-}1.747 & 0.956 & 0.165 & 0.602 & $-16.3$ & 240 & $-$79.5 & 35.41 & 74.72 \\

T3a & 10.887 & 13.894 & 12.321 & \phantom{0}8.46 & $-1.58$ & 2
& \phantom{-}0.411 & 1.707 & 0.150 & 0.558 & $-16.3$ & 240 & 343.0 & 32.79 & 64.47 \\

T3b & 10.890 & 13.898 & 12.412 & \phantom{0}8.72 & $-1.74$ & 2
& \phantom{-}0.720 & 1.572 & 0.155 & 0.545 & $-16.3$ & 240 & 559.1 & 33.54 & 68.54 \\

T4a & \phantom{0}9.283 & 11.038 & 12.100 & 19.05 & $-4.31$ & 0
& \phantom{-}1.791 & 1.252 & 0.150 & 0.686 & $-16.3$ & 240 & $-$396.8 & 33.41 & 63.17 \\

T4b & \phantom{0}9.246 & 11.037 & 11.913 & 18.54 & $-4.38$ & 0
& \phantom{-}1.965 & 1.060 & 0.155 & 0.676 & $-16.3$ & 240 & $-$379.2 & 34.12 & 66.91 \\

T4c & \phantom{0}9.212 & 11.036 & 11.729 & 18.07 & $-4.43$ & 0
& \phantom{-}2.115 & 0.891 & 0.160 & 0.666 & $-16.3$ & 240 & $-$358.0 & 34.86 & 70.75 \\

T4d & \phantom{0}9.179 & 11.034 & 11.549 & 17.64 & $-4.48$ & 0
& \phantom{-}2.244 & 0.741 & 0.165 & 0.656 & $-16.3$ & 240 & $-$332.5 & 35.61 & 74.68 \\

T5a & 10.063 & 12.514 & 12.007 & 11.54 & $-2.60$ & 1
& \phantom{-}0.988 & 1.452 & 0.150 & 0.617 & $-16.3$ & 240 & $-$186.4 & 33.09 & 62.94 \\

T5b & 10.037 & 12.512 & 11.901 & 11.40 & $-2.68$ & 1
& \phantom{-}1.208 & 1.298 & 0.155 & 0.605 & $-16.3$ & 240 & $-$104.0 & 33.82 & 66.87 \\

T5c & 10.015 & 12.509 & 11.808 & 11.31 & $-2.76$ & 1
& \phantom{-}1.419 & 1.159 & 0.160 & 0.593 & $-16.3$ & 240 & $-$5.0 & 34.56 & 70.88 \\

T5d & \phantom{0}9.996 & 12.507 & 11.732 & 11.27 & $-2.85$ & 1
& \phantom{-}1.623 & 1.034 & 0.165 & 0.581 & $-16.3$ & 240 & 113.3 & 35.32 & 74.96 \\

T7a & \phantom{0}9.547 & 11.543 & 12.087 & 16.14 & $-3.94$ & 0
& \phantom{-}1.564 & 1.281 & 0.150 & 0.661 & $-16.3$ & 240 & $-$339.7 & 33.30 & 62.89 \\

T7b & \phantom{0}9.512 & 11.541 & 11.915 & 15.73 & $-3.98$ & 0
& \phantom{-}1.741 & 1.109 & 0.155 & 0.650 & $-16.3$ & 240 & $-$308.9 & 34.02 & 66.71 \\

T7c & \phantom{0}9.479 & 11.537 & 11.747 & 15.38 & $-4.02$ & 0
& \phantom{-}1.900 & 0.957 & 0.160 & 0.639 & $-16.3$ & 240 & $-$271.8 & 34.76 & 70.63 \\

T7d & \phantom{0}9.448 & 11.533 & 11.586 & 15.05 & $-4.06$ & 0
& \phantom{-}2.043 & 0.821 & 0.165 & 0.629 & $-16.3$ & 240 & $-$226.6 & 35.51 & 74.64 \\

T8a & \phantom{0}9.990 & 12.372 & 12.044 & 12.30 & $-3.03$ & 0.5
& \phantom{-}1.099 & 1.413 & 0.150 & 0.621 & $-16.3$ & 240 & $-$194.5 & 33.10 & 62.79 \\

T8b & \phantom{0}9.962 & 12.368 & 11.920 & 12.10 & $-3.09$ & 0.5
& \phantom{-}1.305 & 1.261 & 0.155 & 0.609 & $-16.3$ & 240 & $-$120.8 & 33.83 & 66.75 \\

T8c & \phantom{0}9.937 & 12.365 & 11.808 & 11.95 & $-3.15$ & 0.5
& \phantom{-}1.498 & 1.124 & 0.160 & 0.597 & $-16.3$ & 240 & $-$30.9 & 34.58 & 70.80 \\

T8d & \phantom{0}9.914 & 12.361 & 11.709 & 11.84 & $-3.22$ & 0.5
& \phantom{-}1.684 & 1.001 & 0.165 & 0.585 & $-16.3$ & 240 & 77.3 & 35.34 & 74.94 \\

T9a & 10.535 & 13.313 & 12.149 & \phantom{0}9.67 & $-2.32$ & 1
& \phantom{-}0.610 & 1.607 & 0.150 & 0.576 & $-16.3$ & 240 & 160.0 & 32.85 & 63.53 \\

T9b & 10.521 & 13.312 & 12.128 & \phantom{0}9.72 & $-2.42$ & 1
& \phantom{-}0.865 & 1.472 & 0.155 & 0.562 & $-16.3$ & 240 & 329.5 & 33.59 & 67.70 \\

T9c & 10.512 & 13.311 & 12.134 & \phantom{0}9.82 & $-2.53$ & 1
& \phantom{-}1.111 & 1.351 & 0.160 & 0.549 & $-16.3$ & 240 & 535.2 & 34.34 & 71.95 \\

\hline
Y1 & \phantom{0}9.523 & 11.470 & 12.159 & 16.71 & $-3.97$ & 0
& \phantom{-}1.521 & 1.351 & 0.148 & 0.669 & $-16.3$ & 240 & $-$359.7 & 33.03 & 61.43 \\

Y2 & 10.131 & 12.615 & 12.060 & 11.31 & $-2.52$ & 1
& \phantom{-}0.848 & 1.534 & 0.148 & 0.617 & $-16.3$ & 240 & $-$185.9 & 32.77 & 61.40 \\

Y3 & 10.320 & 12.950 & 12.074 & 10.28 & $-2.22$ & 1.25
& \phantom{-}0.672 & 1.596 & 0.148 & 0.602 & $-16.3$ & 240 & $-$92.2 & 32.69 & 61.60 \\

\hline
T10 & \phantom{0}9.718 & 11.798 & 12.250 & 15.31 & $-3.71$ & 0
& \phantom{-}1.237 & 1.491 & 0.145 & 0.659 & $-16.3$ & 240 & $-$332.6 & 32.53 & 58.99 \\

T11 & 10.203 & 12.697 & 12.180 & 11.49 & $-2.81$ & 0.5
& \phantom{-}0.701 & 1.633 & 0.145 & 0.617 & $-16.3$ & 240 & $-$169.0 & 32.29 & 58.90 \\

T12 & 10.809 & 13.731 & 12.280 & \phantom{0}8.97 & $-2.13$ & 1
& \phantom{-}0.126 & 1.853 & 0.145 & 0.568 & $-16.3$ & 240 & \phantom{-}247.9 & 31.98 & 60.06 \\

T13 & \phantom{0}9.852 & 12.048 & 12.239 & 14.18 & $-3.55$ & 0
& \phantom{-}1.107 & 1.516 & 0.145 & 0.647 & $-16.3$ & 240 & $-$292.4 & 32.46 & 58.85 \\

T14 & 10.389 & 13.021 & 12.194 & 10.60 & $-2.67$ & 0.5
& \phantom{-}0.515 & 1.697 & 0.145 & 0.600 & $-16.3$ & 240 & $-$54.2 & 32.18 & 59.01 \\

T15 & 11.087 & 14.162 & 12.428 & \phantom{0}8.48 & $-2.07$ & 1
& $-$0.104 & 1.967 & 0.145 & 0.545 & $-16.3$ & 240 & \phantom{-}570.1 & 31.82 & 61.20 \\

\end{longtable}
\endgroup
%\twocolumngrid

%\clearpage
%\onecolumngrid
\section{Sensitivity to the saturation density}
\label{app:n0}

The T1--T9 models use $n_0=0.148\,\mathrm{fm^{-3}}$.  We generated controls
up to $0.165\,\mathrm{fm^{-3}}$, readjusting $m^\ast$ to preserve
$M_{N,\mathrm{TOV}}$.  The lower limit $m^\ast\geq0.54$ truncates the T3
and T9 sequences at $n_0=0.155$ and $0.160\,\mathrm{fm^{-3}}$,
respectively, and precludes additional T6 controls.  Table~\ref{tab:n0dep}
lists the individual controls together with the corresponding maximum masses.

Over these ranges, $M_{Y,\mathrm{TOV}}$ changes by at most
$0.012\,M_\odot$ and $M_{Y,707}$ by at most $0.017\,M_\odot$.  Both are
smaller than the changes along the $\zeta$ sequences used to locate the
boundary, so the fixed choice of $n_0$ does not affect the constraint at the
resolution of the present construction.

\begin{table*}[h]
\caption{Maximum masses of the fixed-$M_{N,\mathrm{TOV}}$ controls,
grouped by nucleonic baseline mass.
Suffixes $a$--$d$ denote $n_0=0.150$, $0.155$, $0.160$, and
$0.165\,\mathrm{fm^{-3}}$, respectively; the unsuffixed models have
$n_0=0.148\,\mathrm{fm^{-3}}$.
The saturation density $n_0$ is in fm$^{-3}$, $m^\ast$ is dimensionless,
and the masses are in $M_\odot$.
Within each group $M_{N,\mathrm{TOV}}$ is held at the value given in the
column heading.}
\label{tab:n0dep}
\begin{ruledtabular}
\footnotesize
\renewcommand{\arraystretch}{1.25}
\begin{tabular}{ccccc|ccccc|ccccc}
\multicolumn{5}{c|}{$M_{N,\mathrm{TOV}}=2.350\,M_\odot$} &
\multicolumn{5}{c|}{$M_{N,\mathrm{TOV}}=2.400\,M_\odot$} &
\multicolumn{5}{c}{$M_{N,\mathrm{TOV}}=2.500\,M_\odot$} \\
Model & $n_0$ & $m^\ast$ & $M_{Y,\mathrm{TOV}}$ & $M_{Y,707}$ &
Model & $n_0$ & $m^\ast$ & $M_{Y,\mathrm{TOV}}$ & $M_{Y,707}$ &
Model & $n_0$ & $m^\ast$ & $M_{Y,\mathrm{TOV}}$ & $M_{Y,707}$ \\
\hline
T1  & 0.148 & 0.701 & 1.950 & 2.007 & T4  & 0.148 & 0.690 & 1.993 & 2.052 & T7  & 0.148 & 0.665 & 2.079 & 2.144 \\
T1a & 0.150 & 0.698 & 1.949 & 2.005 & T4a & 0.150 & 0.686 & 1.991 & 2.050 & T7a & 0.150 & 0.661 & 2.078 & 2.142 \\
T1b & 0.155 & 0.688 & 1.945 & 2.000 & T4b & 0.155 & 0.676 & 1.988 & 2.045 & T7b & 0.155 & 0.650 & 2.077 & 2.140 \\
T1c & 0.160 & 0.678 & 1.941 & 1.995 & T4c & 0.160 & 0.666 & 1.985 & 2.042 & T7c & 0.160 & 0.639 & 2.075 & 2.137 \\
T1d & 0.165 & 0.669 & 1.939 & 1.991 & T4d & 0.165 & 0.656 & 1.983 & 2.038 & T7d & 0.165 & 0.629 & 2.074 & 2.135 \\
\hline
T2  & 0.148 & 0.641 & 1.990 & 2.062 & T5  & 0.148 & 0.622 & 2.038 & 2.115 & T8  & 0.148 & 0.626 & 2.109 & 2.185 \\
T2a & 0.150 & 0.636 & 1.989 & 2.060 & T5a & 0.150 & 0.617 & 2.038 & 2.114 & T8a & 0.150 & 0.621 & 2.109 & 2.184 \\
T2b & 0.155 & 0.625 & 1.987 & 2.057 & T5b & 0.155 & 0.605 & 2.036 & 2.111 & T8b & 0.155 & 0.609 & 2.108 & 2.182 \\
T2c & 0.160 & 0.614 & 1.985 & 2.054 & T5c & 0.160 & 0.593 & 2.035 & 2.109 & T8c & 0.160 & 0.597 & 2.107 & 2.180 \\
T2d & 0.165 & 0.602 & 1.984 & 2.052 & T5d & 0.165 & 0.581 & 2.034 & 2.107 & T8d & 0.165 & 0.585 & 2.106 & 2.178 \\
\hline
T3  & 0.148 & 0.563 & 2.030 & 2.118 & T6  & 0.148 & 0.534 & 2.082 & 2.176 & T9  & 0.148 & 0.581 & 2.138 & 2.225 \\
T3a & 0.150 & 0.558 & 2.029 & 2.116 &     &       &       &       &       & T9a & 0.150 & 0.576 & 2.137 & 2.224 \\
T3b & 0.155 & 0.545 & 2.027 & 2.113 &     &       &       &       &       & T9b & 0.155 & 0.562 & 2.136 & 2.222 \\
    &       &       &       &       &     &       &       &       &       & T9c & 0.160 & 0.549 & 2.135 & 2.220 \\
\end{tabular}
\end{ruledtabular}
\end{table*}
\clearpage
\twocolumngrid

\section{Boundary validation: models Y1--Y3}
\label{app:Yvalidation}

\begin{figure}[!htbp]
    \centering
    \includegraphics[width=0.95\linewidth]{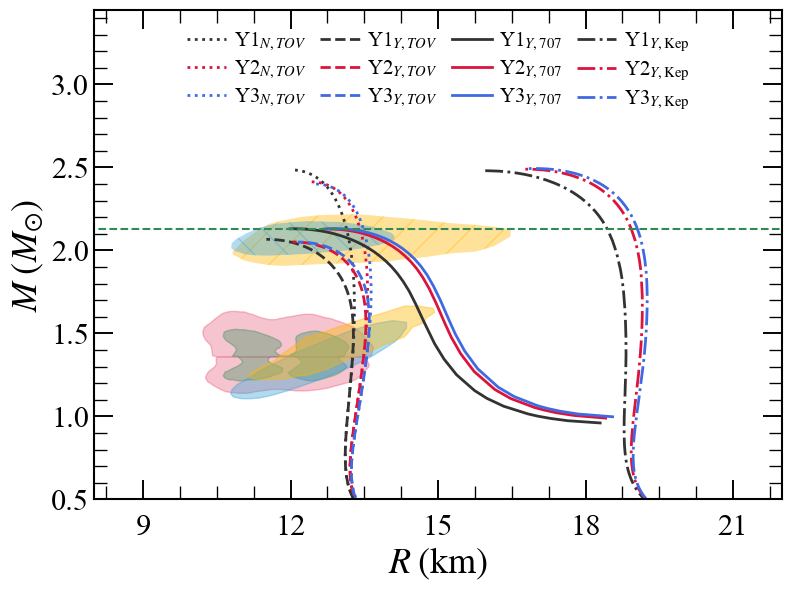}
    \caption{Mass--radius relations for Y1--Y3, constructed at three
    threshold points of the mass-constraint boundary.
    Curve styles and observational regions follow
    Fig.~\ref{fig:M1-10}.}
    \label{fig:Ymodels}
\end{figure}

The Y1--Y3 EOSs were constructed at three points inferred from the
interpolated boundary for $M_{Y,707}=2.13\,M_\odot$.
Their directly computed values, $M_{Y,707}=2.129$--$2.131\,M_\odot$,
confirm that the inferred points are realized by explicit EOSs.
Figure~\ref{fig:Ymodels} shows the corresponding mass--radius sequences;
the model inputs and maximum masses are given in
Tables~\ref{tab:supp-couplings} and \ref{tab:Ymass}.

\begin{table}[!htbp]
\caption{Maximum masses and central hyperon fractions for Y1--Y3.
Notation and units follow Table~\ref{tab:T1-9Mmax}.}
\label{tab:Ymass}
\begin{ruledtabular}
\renewcommand{\arraystretch}{1.25}
\begin{tabular}{cccccc}
Model & $M_{N,\mathrm{TOV}}$ & $M_{Y,\mathrm{TOV}}$ & $M_{Y,707}$ & $M_{Y,\mathrm{Kep}}$ & $y_Y(n_c^{\mathrm{max}})$ \\
\hline
Y1 & 2.485 & 2.067 & 2.130 & 2.480 & 0.526 \\
Y2 & 2.413 & 2.051 & 2.129 & 2.489 & 0.455 \\
Y3 & 2.400 & 2.050 & 2.131 & 2.493 & 0.443 \\
\end{tabular}
\end{ruledtabular}
\end{table}

\section{Constraints from the $1\sigma$ mass threshold}
\label{app:sigma1}

To quantify the dependence on the adopted observational threshold, we repeat
the construction at the $1\sigma$ lower edge of the PSR~J0952$-$0607
measurement, $M_{Y,707}=2.24\,M_\odot$.  The $2\sigma$ value remains the
fiducial choice for the reasons given in Sec.~\ref{sec:results}.  Because the
$707\,\mathrm{Hz}$ rotation is included explicitly, the threshold is compared
directly with the maximum gravitational mass along the rotating sequence.

\begin{figure}[!t]
    \centering
    \includegraphics[width=0.92\linewidth]{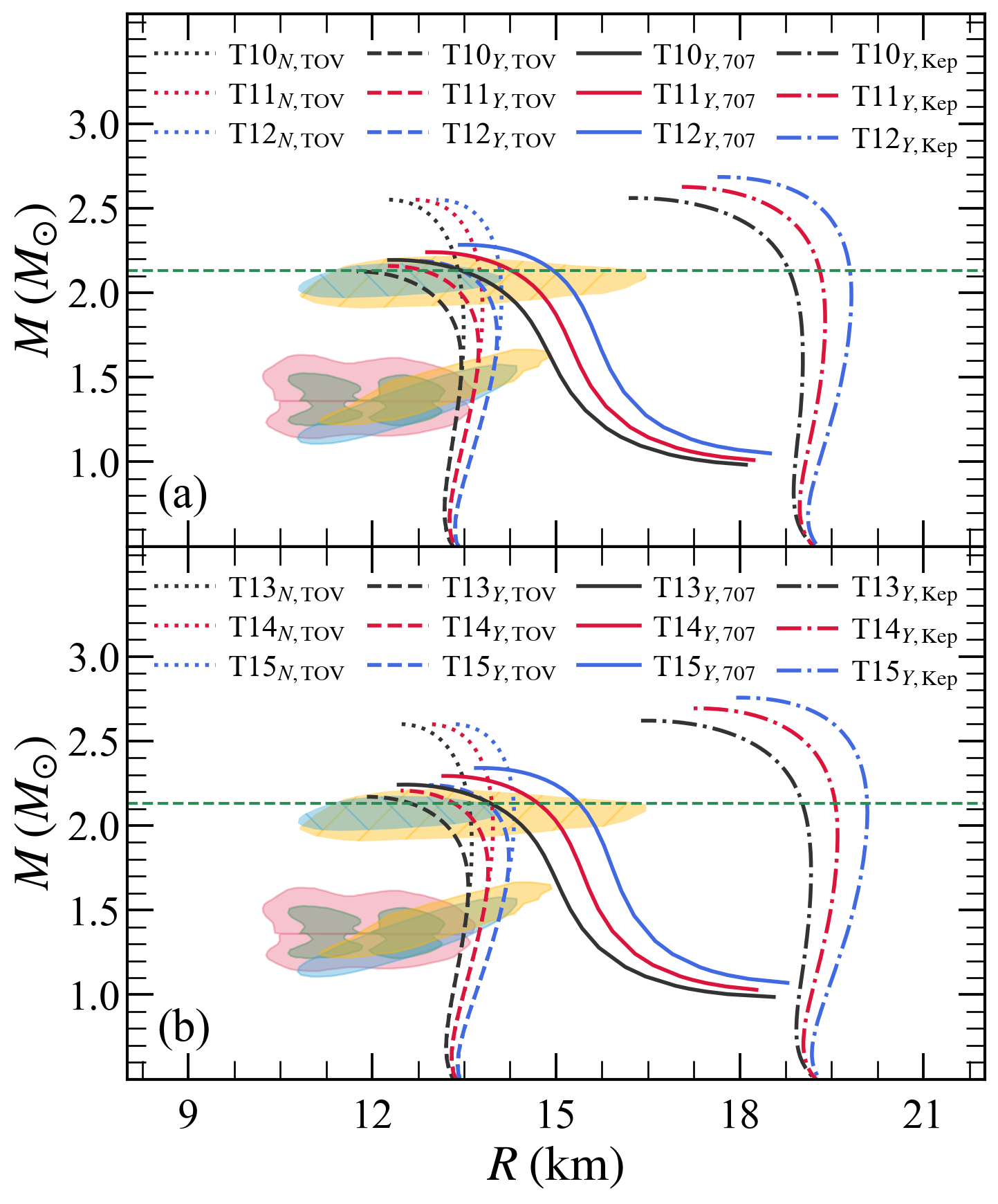}
    \caption{Mass--radius relations for T10--T15:
    (a) T10--T12 with $M_{N,\mathrm{TOV}}=2.55\,M_\odot$ and
    (b) T13--T15 with $2.60\,M_\odot$.
    Curve styles and observational regions follow
    Fig.~\ref{fig:T1-9}.}
    \label{fig:T10-15}
\end{figure}

The T10--T15 sets supply the higher nucleonic baselines $2.55$ and
$2.60\,M_\odot$ required to bracket this tighter threshold.  They use the
same construction as T7--T9 at $n_0=0.145\,\mathrm{fm^{-3}}$ and
$\zeta=0$, $0.5\times10^{-3}$, and $10^{-3}$; their inputs and stellar
masses are listed in Tables~\ref{tab:supp-couplings} and
\ref{tab:T1015mass}.  Figure~\ref{fig:T10-15} shows the resulting
mass--radius sequences.  The unrounded T11 value,
$M_{Y,707}=2.2399\,M_\odot$, lies just below the threshold, whereas T12--T15
lie above it.

% tall 2-panel float: "!" is needed to override \topfraction

\begin{table}[!htbp]
\caption{Maximum masses and central hyperon fractions for T10--T15.
Notation and units follow Table~\ref{tab:T1-9Mmax}.}
\label{tab:T1015mass}
\begin{ruledtabular}
\renewcommand{\arraystretch}{1.25}
\begin{tabular}{cccccc}
Model & $M_{N,\mathrm{TOV}}$ & $M_{Y,\mathrm{TOV}}$ & $M_{Y,707}$ & $M_{Y,\mathrm{Kep}}$ & $y_Y(n_c^{\mathrm{max}})$ \\
\hline
T10 & 2.550 & 2.130 & 2.190 & 2.560 & 0.528 \\
T11 & 2.550 & 2.160 & 2.240 & 2.630 & 0.484 \\
T12 & 2.550 & 2.190 & 2.280 & 2.680 & 0.454 \\
T13 & 2.600 & 2.170 & 2.240 & 2.620 & 0.532 \\
T14 & 2.600 & 2.210 & 2.290 & 2.690 & 0.486 \\
T15 & 2.600 & 2.240 & 2.340 & 2.760 & 0.455 \\
\end{tabular}
\end{ruledtabular}
\end{table}

\begin{figure*}[t]
    \centering
    \includegraphics[width=\textwidth]{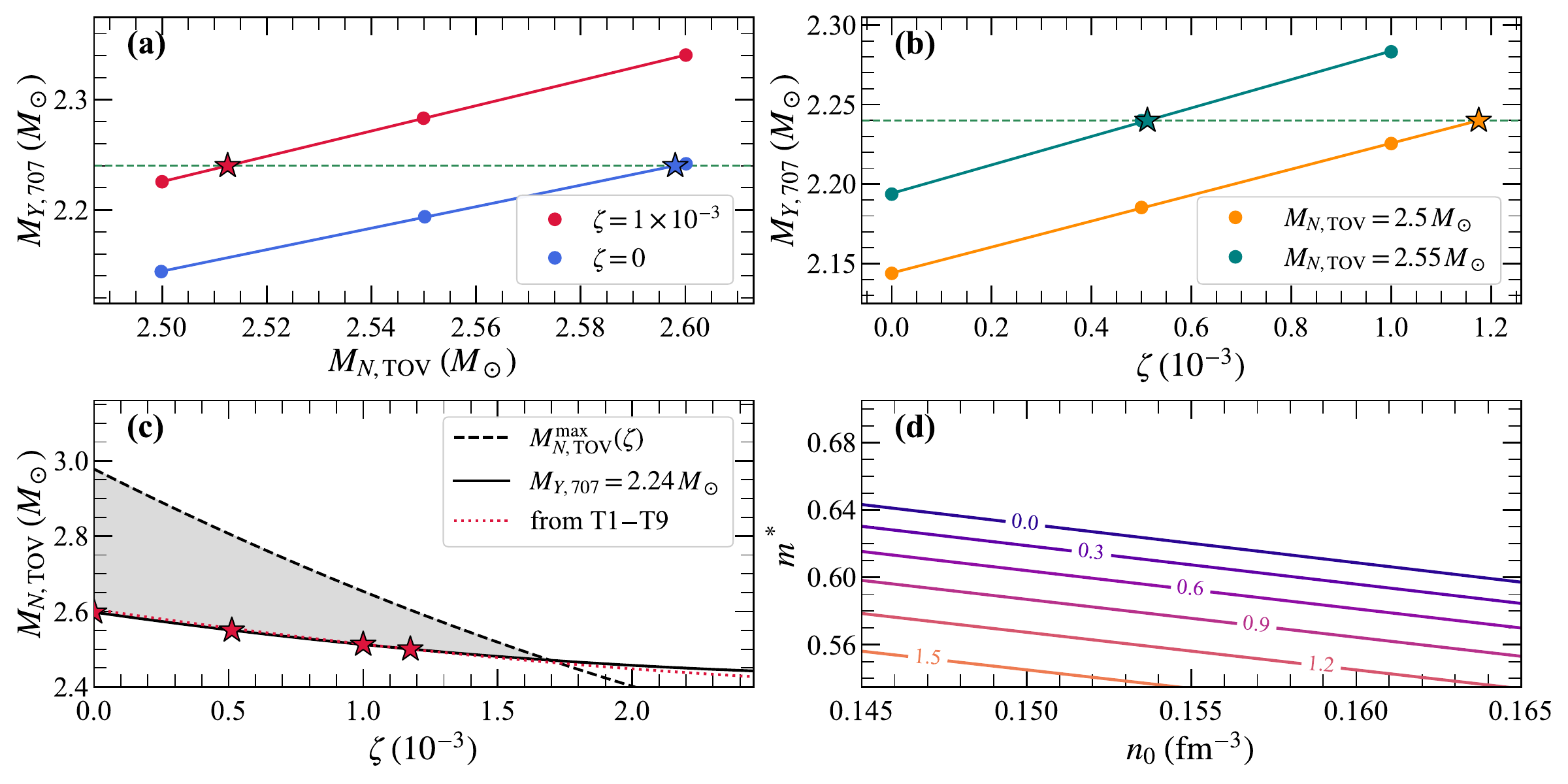}
    \caption{Constraints obtained from the lower mass threshold
    $M_{Y,707}=2.24\,M_\odot$ using T7--T15.
    Panels (a) and (b) follow Fig.~\ref{fig:NMrange-intersections}, and
    panels (c) and (d) follow Fig.~\ref{fig:NMrange-constraints}.
    The fixed baselines in panel (b) are
    $M_{N,\mathrm{TOV}}=2.50$ and $2.55\,M_\odot$; the dotted curve in
    panel (c) is the boundary obtained from T1--T9.}
    \label{fig:NMrange224}
\end{figure*}

Figure~\ref{fig:NMrange224} repeats the construction of
Figs.~\ref{fig:NMrange-intersections} and
\ref{fig:NMrange-constraints} at $M_{Y,707}=2.24\,M_\odot$.  T7--T15 provide
baselines of $2.50$, $2.55$, and $2.60\,M_\odot$ at three values of
$\zeta$.  Intersections of the fixed-$\zeta$ and fixed-baseline fits locate
four threshold pairs; adding the intermediate sequences moves the boundary by
less than $0.002\,M_\odot$.  Their quadratic fit is
\begin{equation}
M_{N,\mathrm{TOV}}(\zeta)
= 1.537\times10^{4}\,\zeta^2 - 101.2\,\zeta + 2.598 ,
\end{equation}
which meets $M_{N,\mathrm{TOV}}^{\max}(\zeta)$ at
\begin{equation*}
(\zeta,\,M_{N,\mathrm{TOV}})\approx(1.70\times10^{-3},\;2.471\,M_\odot).
\end{equation*}
Relative to the fiducial result, the tighter threshold raises the nucleonic
baseline at the closing point from $2.349$ to $2.471\,M_\odot$ and lowers
the corresponding $\zeta$ from $2.25\times10^{-3}$ to
$1.70\times10^{-3}$.

Using T1--T9 alone gives
$M_{N,\mathrm{TOV}}(\zeta)=1.323\times10^{4}\zeta^2-105.3\,\zeta+2.606$ and
$(\zeta,M_{N,\mathrm{TOV}})=(1.73\times10^{-3},2.463\,M_\odot)$, shown by
the dotted curve in Fig.~\ref{fig:NMrange224}(c).  The two boundaries agree
within $0.008\,M_\odot$, but T1--T9 provide only one fixed-baseline sequence
that crosses the tighter threshold within the sampled range.  T10--T15 supply
the second crossing and thereby replace that extrapolation with a bracketed
interpolation without materially changing the result.

\newpage
\section{High-density limit of the hyperonic EOS}
\label{app:validity}

The nucleonic mean-field equations can be continued to higher density
without obstructing the construction of the EOS.  Once hyperons are
included, however, a finite density is reached above which the coupled
equations no longer admit a physical solution with $m_N^\ast>0$, even
though the corresponding nucleonic EOS remains well defined.  
The loss of a positive effective mass in multicomponent relativistic mean-field
matter has been reported before: Knorren et al.\ gave an asymptotic argument for
baryons of unequal masses whose $\sigma$ couplings are not proportional to those
masses~\cite{Knorren1995}, while hyperonic calculations have either been
continued past this point with $|m_N^\ast|$~\cite{Schaffner1996} or terminated
where it occurs~\cite{Hofmann2001}.  This appendix derives a finite-density
criterion for the end of the $m_N^\ast>0$ branch in the present nonlinear
functional, determines the critical density, and compares it with the densities
reached inside the stars.
\clearpage
\onecolumngrid
{\centering
\includegraphics[width=\textwidth]{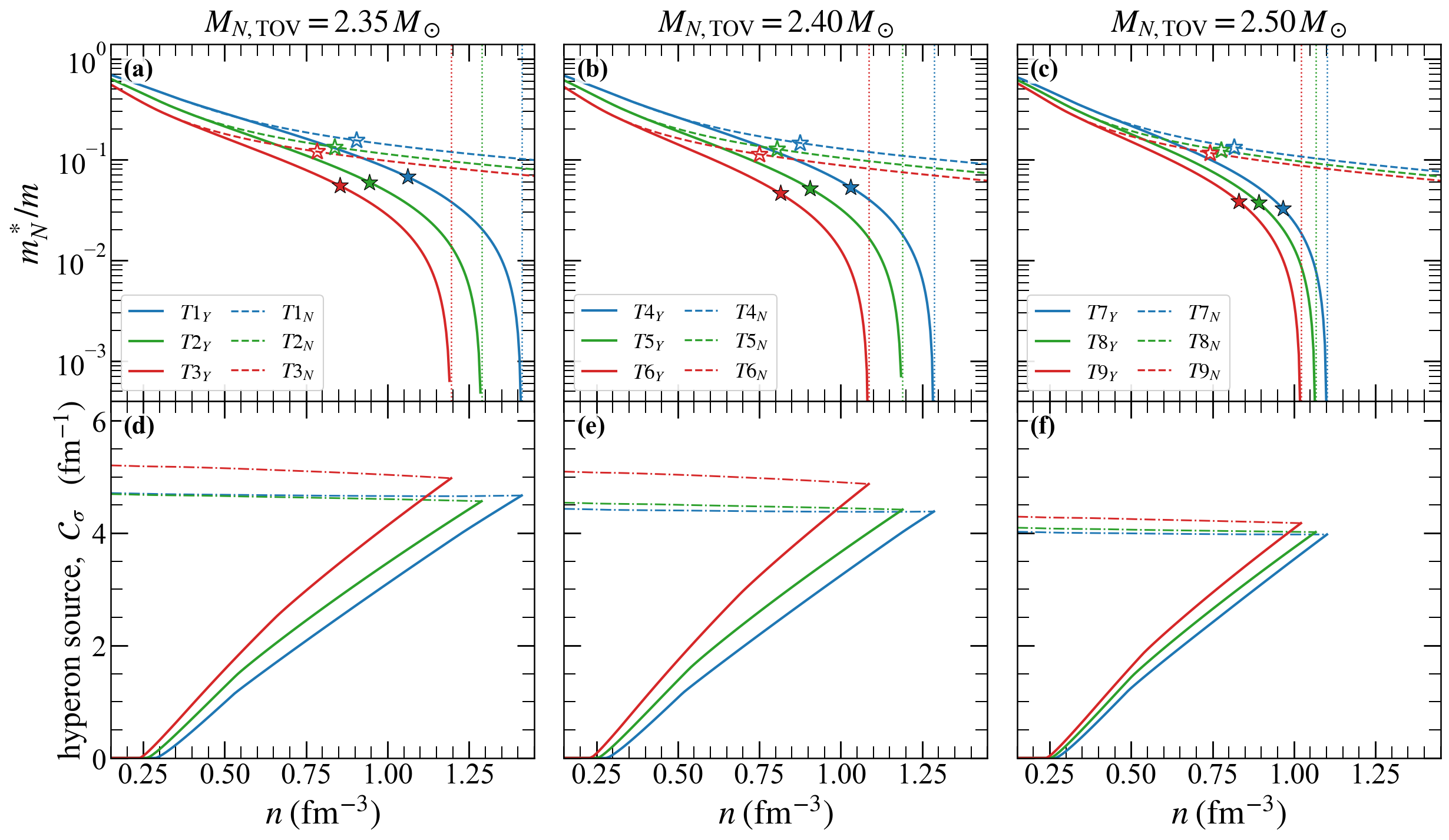}\par}
\captionof{figure}{High-density limit of the hyperonic EOS for T1--T9.
    Top row: dimensionless nucleon Dirac mass $m_N^\ast/m$ for
    hyperonic (solid) and nucleonic (dashed) matter.
    Dotted vertical lines mark $n_{\mathrm{crit}}$, and filled and open
    stars mark the central density of the hyperonic and nucleonic
    nonrotating maximum-mass configurations.
    Bottom row: hyperon scalar source
    $(g_\sigma/m_\sigma^2)\sum_Yg_{\sigma Y}n_s^Y$ (solid) and the
    threshold $\mathcal{C}_\sigma(n)$ (dash-dotted), evaluated in the
    $m_N^\ast\rightarrow0$ limit; their common endpoints define
    $n_{\mathrm{crit}}$.
    Columns correspond to nucleonic maximum masses
    $M_{N,\mathrm{TOV}}=2.35$, $2.40$, and $2.50\,M_\odot$.}
\label{fig:validity}
\twocolumngrid

The distinction follows from the $\sigma$-field equation,
Eq.~\eqref{eq:sigma_field}.  In nucleonic matter, the nucleon scalar source
decreases together with $m_N^\ast$ and vanishes as
$g_\sigma\sigma\rightarrow m$.  The equation therefore retains an
intersection within the physical domain $g_\sigma\sigma<m$.  With
hyperons, their Dirac masses instead approach
$m_Y-(g_{\sigma Y}/g_\sigma)m$ and remain positive.  The hyperon scalar
source consequently persists and grows as the hyperon populations
increase.  At sufficiently high density it exceeds the boundary value
that the $\sigma$ field can support, leaving no intersection within the
physical domain.

This limitation is obtained by evaluating Eq.~\eqref{eq:sigma_field} at
$g_\sigma\sigma=m$, where the nucleon scalar source vanishes.  The
field-side boundary is
\begin{align}
\mathcal{C}_\sigma(n)
\equiv{}&
m+
\left(\frac{g_\sigma}{m_\sigma}\right)^2
\biggl[
\kappa m^2+\lambda m^3
\nonumber\\
&\qquad
-(g_\rho\rho_{03})^2
\bigl(\Lambda_{s1}+2\Lambda_{s2}m\bigr)
\biggr],
\label{eq:sigma_threshold}
\end{align}
where $\rho_{03}$ retains the density dependence of the mixed-interaction
term.  A solution with $m_N^\ast>0$ exists while
\begin{equation}
\left.
\frac{g_\sigma}{m_\sigma^2}
\sum_Y g_{\sigma Y}n_s^Y
\right|_{g_\sigma\sigma\rightarrow m}
<
\left.
\mathcal{C}_\sigma(n)
\right|_{g_\sigma\sigma\rightarrow m}.
\label{eq:hyperon_critical}
\end{equation}
Both sides of Eq.~\eqref{eq:hyperon_critical} are evaluated in an
auxiliary system with $m_N^\ast\rightarrow0$, not by extending the
equilibrium EOS.  Equality defines $n_{\mathrm{crit}}$, where the
lower-panel curves terminate.  Above $n_{\mathrm{crit}}$ the hyperon
source exceeds $\mathcal{C}_\sigma(n)$, so satisfying the $\sigma$-field
equation would require $g_\sigma\sigma>m$, or $m_N^\ast<0$.  Hence no
physical equilibrium solution exists within the domain $m_N^\ast>0$.

Figure~\ref{fig:validity} compares $n_{\mathrm{crit}}$ with the central
density $n_c$ of the maximum-mass hyperonic configuration.  In the
closest case, T7, $n_c=0.966\,\mathrm{fm^{-3}}$ lies $12.3\%$ below
$n_{\mathrm{crit}}=1.101\,\mathrm{fm^{-3}}$.  T3 has the largest
separation, $28.6\%$, with $n_c=0.853\,\mathrm{fm^{-3}}$ and
$n_{\mathrm{crit}}=1.195\,\mathrm{fm^{-3}}$.

Within each fixed-$M_{N,\mathrm{TOV}}$ group, $n_0$ is also fixed, so a
larger $\zeta$ requires a smaller $m^\ast$.  The nucleon Dirac mass then
starts lower at saturation and reaches zero at a lower density, shifting
$n_{\mathrm{crit}}$ downward from T1 to T3, T4 to T6, and T7 to T9.
The maximum-mass central density $n_c$ also decreases.  These shifts
preserve $n_c<n_{\mathrm{crit}}$, while the fractional separation grows
from $24.9\%$ to $28.6\%$, $19.8\%$ to $25.0\%$, and $12.3\%$ to
$18.7\%$ across the three groups.  Thus larger $\zeta$ lowers the
intrinsic EOS endpoint, but the maximum-mass configuration is reached at
lower density as well.

\end{document}